\documentclass{iopart}
\pdfoutput=1

\usepackage{iopams}

\expandafter\let\csname equation*\endcsname\relax

\expandafter\let\csname endequation*\endcsname\relax

\usepackage{float}
\usepackage{amsmath}
\usepackage{subfig}
\usepackage{pifont}

\makeatletter
\let\@afterindenttrue\@afterindentfalse
\makeatother
\usepackage{scalefnt}
\usepackage{graphicx}
\usepackage[utf8]{inputenc}
\usepackage{tikz}
\usepackage[a4paper, margin=3cm]{geometry} 
\usepackage[]{algorithm2e}

\usepackage{hyperref}

\definecolor{mygreen}{rgb}{0.0,0.6,0.45}

\newcommand{\Nc}{n_\mathrm{c}}
\newcommand{\tmax}{t_{\rm obs}}
\newcommand{\CC}{\mathcal{C}}
\newcommand{\KK}{\mathcal{K}}
\newcommand{\fs}{\hat{k}} 
\newcommand{\pend}{p_{\rm end}}
\newcommand{\pave}{p_{\rm ave}}

\newcommand{\XX}{\mathcal{X}}

\graphicspath{{./Figures/}}

\begin{document}

\title{Efficient characterisation of large deviations using population dynamics}

\author{Tobias Brewer$^1$, Stephen R. Clark$^{1,2}$  Russell Bradford${^3}$, and Robert L. Jack$^{1,4,5}$, }
\address{$^1$ Department of Physics, University of Bath, Bath BA2 7AY, United Kingdom}
\address{$^2$ Max Planck Institute for the Structure and Dynamics of Matter,
University of Hamburg CFEL, Hamburg, Germany}
\address{$^3$ Department of Computer Science, University of Bath, Bath BA2 7AY, United Kingdom}
\address{$^4$ Department of Applied Mathematics and Theoretical Physics, University of Cambridge, Wilberforce Road, Cambridge CB3 0WA, United Kingdom}
\address{$^5$ Department of Chemistry, University of Cambridge, Lensfield Road, Cambridge CB2 1EW, United Kingdom}

\begin{abstract}
We consider population dynamics as implemented by the cloning algorithm for analysis of large deviations of time-averaged quantities. We use the simple symmetric exclusion process with periodic boundary conditions as a prototypical example and investigate the convergence of the results with respect to the algorithmic parameters, focussing on the dynamical phase transition between homogeneous and inhomogeneous states, where convergence is relatively difficult to achieve.  We discuss how the performance of the algorithm can be optimised, and how it can be efficiently exploited on parallel computing platforms. 
\end{abstract}

\section{Introduction}
Large deviation theory~\cite{touchette2009} is concerned with the probabilities of rare events in random processes.  Despite their scarcity, such events can have dramatic consequences, especially if one considers the behaviour of systems on long time scales, as in geology~\cite{gretener1967significance} or climate science~\cite{easterling2000climate}. 
In theoretical physics, a number of recent studies (for example~\cite{derrida07,garrahan07,hedges09,speck12,weber2013emergence}) have concentrated on rare events in which a system does not behave ergodically -- that is, time-averaged observable quantities have non-typical values, even when measured over long time periods. 
While, it is challenging to capture rare events in experiments (see however~\cite{pinch17}), there are a range of computational approaches for the characterisation of rare events~\cite{bolhuis02,allen05,giardina06,huber96,zhang10,bouchet16,johnson2015capturing,PhysRevE.82.036702} -- these are vital since analytical calculations are usually possible only in very simple systems~\cite{derrida07}.
Computational analyses of large deviations using these methods have lead to new insights in glassy materials \cite{jack2006space, garrahan2009first}, protein-folding \cite{weber2013emergence,weber2014dynamical,mey2014rare} and integrable systems \cite{tailleur2007probing}.  In particular, analysis of the associated rare-event mechanisms can reveal properties of metastable or unusual dynamical states that can aid understanding of the typical behaviour.

When considering the large deviations of time-averaged quantities, the two dominant
computational methods are transition path sampling~\cite{bolhuis02} and a cloning (or population dynamics) algorithm~\cite{giardina06}.  
This article investigates properties of the cloning method, applied to the symmetric simple exclusion process (SSEP).  This is a prototypical interacting-particle system, for which analytical results are available~\cite{derrida07,appert08,lecomte2012inactive}, allowing a direct comparison between theory and simulation.  While the model is very simple, it exhibits a range of surprising rare-event phenomena, including events where many particles assemble into a large (macroscopic) cluster, as well as hyperuniform states~\cite{thompson2015dynamical}, where density fluctuations are strongly suppressed.  These regimes of behaviour are separated by dynamical phase transitions~\cite{bodineau2005distribution,bodineau2007cumulants, bodineau2012finite}: at these points, numerical calculations become challenging and require analysis of large system sizes and long time scales.

The cloning algorithm was introduced more than ten years ago~\cite{giardina06,lecomte2007numerical}, based on earlier ideas in statistical physics~\cite{grassberger83,grassberger2002go} and quantum mechanics~\cite{anderson75}.  It has been applied to a range of systems ~\cite{tailleur2007probing,pitard11,hurtado14,nemoto2017}.  The method is powerful, but it requires simulations of many copies (``clones'') of the system, and its results are accurate only in the limit where the number of clones tends to infinity.  For finite numbers of clones, there are both systematic and random errors, which have been analysed recently in Refs. ~\cite{nemoto17finite,hidalgo17finite,limmer2017importance}.  The scaling of these errors has been determined using an analytical description of the algorithm and has been verified using a numerical approach to measure how quickly an estimator converges towards its true value~\cite{nemoto17finite}. 
Simple guiding models (or control forces) have been used to improve the efficiency of convergence~\cite{nemoto2016population,nemoto2017,limmer2017importance}.  In previous analyses of the convergence of the algorithm, it was often assumed that the clone population is larger than the total number of states visited by the model~\cite{nemoto17finite,hidalgo17finite}: this is not the case in typical applications so we analyse here the case where the clone population is much smaller than the total number of states.

We have implemented the cloning method on high-performance computing platform, which allows us to investigate large numbers of clones.  We present a detailed analysis of the dynamical phase transition that occurs in the SSEP.  The finite-size scaling properties of this system differ from conventional phase transitions~\cite{appert08,lecomte2012inactive}, and we discuss the physical reasons for this.  In addition, we discuss the systematic errors inherent in the cloning method; we provide practical heuristics as to how the significance of these errors can be assessed; we provide some simple optimisations of the method in order to reduce these errors.  Finally, we discuss our computational implementation, and how this can be optimised to improve performance when the number of clones is large.

The form of this paper is as follows: Section~\ref{sec:largedev} provides background on large deviation theory and the cloning method; Section~\ref{sec:ssep-etc} describes the model and its dynamical phase transition, including some numerical results that allow characterisation of finite-size effects.  Section~\ref{sec:Algorithm} explains in more detail how we apply the method to the SSEP, and how this method is optimised to reduce errors.  After that, Section~\ref{sec:Conv} analyses the convergence of the algorithm with respect to the number of clones, and the (long) time scale associated with the rare events of interest.  Finally, Section~\ref{sec:comput} discusses the computational implementation, and we summarise our conclusions in Section~\ref{sec:conc}.

\section{Large deviation theory and the cloning algorithm}
\label{sec:largedev}

\subsection{Large deviation theory for time averaged-quantities}
\label{sec:timeave}

Consider a physical system described by a Markov process on a discrete set of states (for example, the SSEP). The state of the system at time $t$ is $\CC_t$, the transition rate from state $\CC$ to $\CC'$ is $W(\CC\to\CC')$, and we define the \emph{escape rate} as $r(\CC)=\sum_{\CC'}W(\CC\to\CC')$.    Let $\Theta$ denote a trajectory of the system, during the time interval $[0,\tmax]$, for some \emph{observation time} $\tmax$.  In trajectory $\Theta$, suppose that the configuration changes happen at times $t_1,t_2,\dots t_K$, and let the state of the system just after the $k$th change be $\CC_k$ (also let the initial configuration be $\CC_0$ and define $t_{K+1}=\tmax$ and $t_0=0$). Then, denoting the probability of trajectory $\Theta$ by $P(\Theta)$, one has (see for example~\cite{garrahan2009first}):
\begin{equation}
{P}(\Theta) = \left[ \prod_{k=0}^{K-1}  W(\CC_k \rightarrow \CC_{k+1}) \right] \cdot \exp \left[-\sum_{k=0}^K r(\CC_k) (t_{k+1} - t_{k }) \right] .
\label{equ:P-Theta}
\end{equation}

Now let $A_t$ be a (random) observable quantity that depends on the behaviour of the system during the time-interval $[0,t]$.  For example, in the SSEP, $A_t$ will be the total number of particle hopping events in $[0,t]$, as discussed in~\cite{derrida07,thompson2015dynamical}.  Alternatively, $A_t$ might be a time integral of the form $\int_0^t b(\CC_{t'}) \mathrm{d}t'$, where $b$ is some function that depends on the configuration.  With either of these choices, one expects that for large $t$, the probability distribution of $A_t$ scales as
\begin{equation}
\label{eq:ldp}
\mathrm{Prob}(A_t \approx at) \sim \exp(-\pi(a) t).
\end{equation}
This is an example of a large-deviation principle~\cite{touchette2009,garrahan2009first}. The function $\pi$ is known as the rate function, and satisfies $\pi(a)\geq0$.  Typically, there is a single value $\overline a$ for which $\pi(\overline a)=0$.  In that case, one sees from (\ref{eq:ldp}) that as $t\to\infty$,  the distribution of $a_t=A_t/t$ concentrates on the single value $\overline{a}$, with the probability of any other value being suppressed exponentially in $t$.  Of course, the validity of~(\ref{eq:ldp}) depends on the system of interest and the observable $A$ -- here we consider irreducible Markov processes with finite (discrete) state spaces, for which~(\ref{eq:ldp}) holds for a large set of observables $A_t$: see eg~\cite{garrahan2009first}.

The general aims of rare-event sampling methods in this context are (i)~to estimate the function $\pi(a)$, which gives the probability of rare events (with $a_t\neq \overline{a}$); and (ii)~to characterise the rare events themselves: what trajectories lead to these rare values? To achieve these aims, it is convenient to introduce a \emph{biasing field} -- denoted by $s$ -- which allows access to the relevant rare events. We use $\Theta$ to denote a trajectory of the system, during the time interval [0, $t$], and let $P(\Theta)$ be the probability of trajectory $\Theta$ as defined, for example in~\cite{garrahan2009first}.  The distribution $P(\Theta)$ depends on the  initial condition of the model.  The results of the following large-deviation analysis are independent of the initial condition, but we assume for concreteness that the initial condition is taken from the steady-state probability distribution of the model, so that $P(\Theta)$ corresponds to the steady state. 

Now define a new probability distribution
\begin{equation}
\label{eq:biastraj}
\tilde{P}_t(\Theta,s) = \frac{P(\Theta) \exp[-sA_t(\Theta)]}{Z{(s,t)}} ,
\end{equation}
where $A_t(\Theta)$ is the value of $A_t$ associated with trajectory $\Theta$, and $Z(s,t) = \left< e^{-sA_t} \right>_0$ is a dynamical partition function (normalisation constant). 
Here and throughout, the notation $\langle \cdot \rangle_0$ indicates an average with respect to $P(\Theta)$.  
The average of an observable with respect to $\tilde P_t$ is denoted by
\begin{equation}
\label{eq:Oave}
\left<\mathcal O \right>_s = \int \mathcal O(\Theta) \tilde{P}_t(\Theta, s) d\Theta .
\end{equation}
It is useful to define 
\begin{equation}
\psi(s) = \lim_{t\to\infty} \frac1t \ln Z(s,t) .
\label{equ:psi-z}
\end{equation}
This limit certainly exists if the LDP (\ref{eq:ldp}) holds.

The distribution $\tilde{P}$ is parameterised by the field $s$.  For $s=0$ we recover the original distribution $P$; for $s>0$ trajectories with large values of $A_t$ are suppressed. The advantage of introducing the field $s$ is that averages of the form $\left< \mathcal{\mathcal O} \right>_s $ can often be evaluated by some numerical or analytical method.  In the absence of dynamical phase transitions, one may then obtain the rate function in (\ref{eq:ldp}), as $\pi(a)=\max_s [ -sa - \psi(s) ]$.
Moreover, if the value of $s$ that achieves this maximum is $s^*_a$ then trajectories obtained from the distribution $\tilde{P}(\Theta,s^*_a)$ are representative trajectories associated with the rare event $a_t=a$ discussed above in the large time limit ~\cite{chetrite2015}.  Thus, computational analysis of $\tilde{P}$ can achieve the two aims (i) and (ii) above, to estimate $\pi$ and to characterise the trajectories that realise these rare events.

The function $\psi$ is a scaled cumulant generating function~\cite{garrahan2009first}: one has
\begin{equation} \lim_{t\to\infty}\left< A_t/t \right>_s = -\psi'(s) \label{equ:psi-mean} \end{equation}
where the prime denotes a derivative.
There is also an associated susceptibility (scaled variance)
\begin{equation}
\lim_{t\to\infty}\frac{1}{t} \left [ \left< A_t^2 \right>_s - \left< A_t \right>_s^2 \right ] = \psi''(s) .
\label{equ:psi-var} \end{equation}

\subsection{Modified Dynamics}
\label{subsec:moddyn}

\newcommand{\UU}{U}

Large deviations are hard to analyse computationally because the associated events are rare -- this means that averages such as $\langle {\rm e}^{-sA_t}\rangle_0$ are dominated by trajectories that are not at all typical of the system at equilibrium.  To analyse such events, it is often convenient to modify the dynamics of the model, so that the relevant trajectories become less rare~\cite{giardina06,nemoto2016population,nemoto2017}.  Consider a general system with modified (or ``controlled'') dynamics, and let its path probability distribution be $\hat{P}(\Theta)$, which is analogous to the distribution $P(\Theta)$ for the original system.  For the modifications that we consider, it is possible to relate these two distributions, as
\begin{equation}
\label{eq:modpdef}
P({\Theta}) = \hat{P}({\Theta}) {\rm e}^{-\hat{\UU}_t(\Theta)}
\end{equation}
where $\hat{\UU}_t(\Theta)$ is a weight function that depends on the trajectory $\Theta$.  (A specific example will be considered in Sec.~\ref{subsec:mod-dyn-ssep}, below.)
Hence one has also
\begin{equation}
\label{eq:phattilde}
\tilde{P}_t(\Theta,s) = \frac{1}{Z(s,t)}\hat{P}({\Theta}) {\rm e}^{-[\hat{\UU}_t(\Theta)+sA_t(\Theta)]} .
\end{equation}
The significance of this result is that the distribution $\tilde{P}_t$ can be analysed in many different ways: either directly as in (\ref{eq:biastraj}) or by simultaneously modifying the dynamics of the system, $P(\Theta)\to\hat{P}(\Theta)$, and at the same time modifying the weighting factor as $sA_t(\Theta)\to [sA_t(\Theta)+\hat{\UU}_t(\Theta)]$.  This freedom to modify the dynamics is very useful when designing computational algorithms.
Finally, we define 
\begin{equation}
\label{eq:upsdef}
\Upsilon_t(\Theta) = \exp[-\hat{\UU}_t(\Theta)-sA_t(\Theta)],
\end{equation}
which is the weight that should be associated with trajectory $\Theta$, in order to obtain the distribution (\ref{eq:biastraj}) by importance sampling from the distribution $\hat{P}$. In particular, we have
\begin{equation}
Z(s,t) = \int \Upsilon_t(\Theta) \hat{P}(\Theta) d\Theta .
\label{equ:Z-ups}
\end{equation}

\subsection{Cloning algorithm}
\label{subsec:alg}

The results of this paper use the cloning (or population dynamics) algorithm that was proposed by Giardina, Kurchan and Peliti~\cite{giardina06} as a method for studying large deviations.  As noted above, this method draws on earlier work by Grassberger~\cite{grassberger2002go} as well as Diffusion Quantum Monte Carlo methods~\cite{anderson75,limmer2017importance}.  We outline this algorithm here, further details are provided in Section \ref{sec:Algorithm}, below.
The method is based on simulations of a population of $\Nc$ copies (or clones) of the system, evolving over a total observation time $\tmax$.  The field $s$ is a fixed parameter: to obtain accurate estimates of $\psi(s)$ one requires a limit of large $\Nc$ and $\tmax$.  The dependence of the results of the algorithm on $\Nc$ and $\tmax$ will be discussed in Section \ref{sec:Conv} below.  
In our implementation, the population size is held strictly constant, although modified algorithms with variable populations are also possible~\cite{lecomte2007numerical}.

Within the algorithm, the total time $\tmax$ is split into intervals of length $\Delta t$, so the number of such intervals is $M=\tmax/\Delta t$.  Within each step of the algorithm, each clone evolves independently for a time $\Delta t$.  Then, some clones are deleted and others copied, in order to bias the system towards the rare events of interest (this is a form of importance sampling). 
In the following we refer to these two sub-steps (or stages) as the dynamical stage and the cloning stage of the algorithm.
The full algorithmic step -- dynamics followed by cloning -- is repeated $M$ times.  
The parameter $\Delta t$ can be chosen according to the problem of interest: as $\Nc\to\infty$ (with fixed $\tmax$) then the results are independent of $\Delta t$.  However, $\Delta t$ has significant effects on the accuracy of the results obtained: this is discussed in Section \ref{sec:Interval} below.

We index the time intervals by $\beta=1,2,\dots M$ and define $t_\beta=\beta\Delta t$. Then the cloning method rests on the fact that for any trajectory $\Theta$, one may write
$$
A_{\tmax}(\Theta) = \sum_{\beta=1}^M A^\beta(\Theta)
$$
where $A^\beta(\Theta)$ is the contribution to $A_t(\Theta)$ from the time interval $[t_{\beta-1},t_\beta]$. Similarly 
\begin{equation}
\label{eq:ubdef}
\hat{\UU}_{\tmax}(\Theta) = \sum_{\beta=1}^M \hat{\UU}^{\beta}(\Theta).
\end{equation}
Note that in the definitions of $A^\beta(\Theta), \: \hat{\UU}^{\beta}(\Theta)$ the superscript $\beta$ is an index and should not be confused with an exponent. We also index the clones of the system by $i=1,2,\dots \Nc$.  Then, define a weighting factor for clone $i$ associated with time-interval $\beta$ as
\begin{equation}
\label{eq:upsbdef}
\Upsilon^{\beta}(\Theta_i) = \exp[-\hat{\UU}^{\beta}(\Theta_i)-sA^{\beta}(\Theta_i)].
\end{equation}
where $\Theta_i$ is the trajectory followed by clone $i$.
This weighting factor plays two roles within the algorithm.  First, in the importance sampling step that takes place at time $t_\beta$, the average number of times that clone $i$ is copied is proportional to $\Upsilon^\beta(\Theta_i)$.  Second, based on (\ref{equ:psi-z},\ref{equ:Z-ups}), one may estimate $\psi(s)$ as
\begin{equation}
\hat\psi(s) = \frac{1}{\tmax} \sum_{\beta=1}^M  \ln \left( \frac{1}{\Nc} \sum_{i=1}^{\Nc}  \Upsilon^{\beta}(\Theta_i) \right) .
\label{equ:psihat}
\end{equation}
For a given computation, this estimator is subject to both systematic and random errors.  However, both these errors vanish as $\Nc,\tmax\to\infty$, and the estimator becomes exact.

\subsection{Estimating averages with respect to $\tilde P$.}
\label{subsec:measave}

To estimate averages of the form $\langle \mathcal O\rangle_s$, one starts by considering the population of clones at the final time $\tmax$.  For each clone $i$ in that population, one follows its trajectory backwards in time: this trajectory is denoted by $\hat\Theta_i$.  Note that many members of the final clone population may be descended from a single clone at some earlier time.  Hence, the trajectories $\hat\Theta_i$ are not all independent samples from $\tilde P$.  However, one may estimate the general expectation value (\ref{eq:Oave}) as
\begin{equation}
\label{eq:avedef}
\langle \mathcal O\rangle_s \approx \frac{1}{\Nc} \sum_{i=1}^{\Nc} \mathcal{O}(\hat{\Theta}_i)
\end{equation}
where $\mathcal{O}(\Theta)$ is the value of observable $\cal O$ in trajectory $\Theta$.  The approximate equality becomes exact  \cite{nemoto17finite} as $\Nc\to\infty$. 

The observable $\cal O$ may depend in general on all aspects of the trajectory $\Theta$.  It is also useful to consider a specific class of time-dependent observables: let $F_t$ be a function that depends on the state of the system at time $t$, such as the number of particles on a particular lattice site.  The average of such an observable is $\langle F_{t} \rangle_s$, which may be evaluated for any time $t$ between $0$ and $\tmax$.  For $s=0$, the probability distribution $P$ is time-translation invariant (TTI), which means that $\langle F_{t} \rangle_0$ does not depend on $t$.  However, for $s\neq0$, the average $\langle F_{t} \rangle_s$ depends on $t$: there are initial and final transient regimes for small $t$ and for $t\approx \tmax$, with an intermediate time-translation invariant regime.  That is,
\begin{equation}
\label{eq:TTIave}
\left< F_{t} \right>_s = \left \{
\begin{aligned}
& F_{\mathrm{i}}(t), && t \lesssim  \tau \\
& F_{\mathrm{f}}(\tmax - t), && (\tmax - t) \lesssim \tau \\
& F_{\infty}, && \hbox{otherwise}
\end{aligned}
\right.
\end{equation}
where $\tau$ is a characteristic time scale for the transient regimes, and $F_{\rm i}, F_{\rm f}$ are functions describing the transients, which decay to the asymptotic value $F_\infty$ as their arguments get large~\cite{jack2010large,nemoto2016population}.

It will be useful in the following to consider the probability distribution of $F_{t}$, and not just its mean value.   This distribution is defined as 
$$p_{\tmax}(F,t)=\langle \delta(F-F_{t}) \rangle_s,$$ 
which is the probability (density) to observe the value $F$ for the observable $F_{t}$, given trajectories of length $\tmax$ with distribution $\tilde P$.  To characterise the TTI regime, we define
\begin{equation}
\label{eq:Opave}
\pave(F) = \lim_{\tmax \to \infty} p_{\tmax}(F,\alpha \tmax)
\end{equation}
with $0<\alpha<1$. The result is independent of $\alpha$ because we have both $\alpha \tmax \to \infty$ and $(\tmax-\alpha \tmax) \to \infty$. Hence, $\alpha t$ is a time in the TTI regime, $\alpha t, \: (\tmax-\alpha t)\gg\tau$ and it follows that $F_\infty=\int F \pave(F) \mathrm{d}F$. 
We also define
\begin{equation}
\label{eq:Opend}
\pend(F) = \lim_{\tmax \to \infty} p_{\tmax}(F,\tmax)
\end{equation}
which is the distribution of $F$ at time $t=\tmax$.  

Note that we have focussed here on instantaneous observables: $F_{t}$ depends only on the configuration of the system at time $t$.  However, the definitions of $\pave,\:\pend$ can be straightforwardly generalised to observables that depend on the trajectory of the system, within a finite time window $[t,t+\Delta t]$.

Within the cloning algorithm, the relevance of $\pave$ and $\pend$ is that the clone population just after the importance sampling step is distributed as $\pend$.  On the other hand, the distribution $\pave$ characterises the ``ancestral population'': this is the distribution that one obtains by constructing the trajectories $\hat\Theta$ from the current population by following their histories backwards in time.  For a detailed discussion see Ref~\cite{nemoto2016population}.

\section{Dynamical phase transition in the symmetric simple  exclusion process}
\label{sec:ssep-etc}

In the following, we apply the cloning algorithm to a model system: the symmetric simple exclusion process (SSEP).

\subsection{Model and choice of dynamical observable}

\begin{figure*}
\centering
\includegraphics{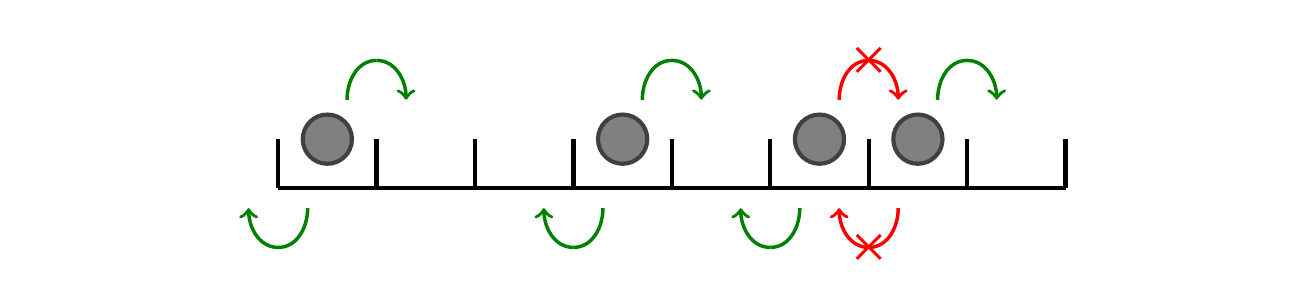}
\caption{Illustration of the SSEP on a one-dimensional lattice of 8 sites with periodic boundary conditions and $N=4$ particles.  Each hop is attempted with rate 1; all possibilities for the attempted hop are indicated with arrows.
}
\label{fig:PT}
\end{figure*}

We consider a one-dimensional lattice of $L$ sites with periodic boundaries. The lattice is occupied by $N$ particles with at most one particle per site. Each particle attempts to hop with rate 1 to each of its neighbouring sites as in Figure \ref{fig:PT}. The attempted hop is successful if the neighbouring site is unoccupied.  Let the occupancy of site $i$ be $n_i$. The model obeys detailed balance which means that in its equilibrium (steady) state, the occupancy of each site is independent: $n_i=1$ with probability $\rho=N/L$ and $n_i=0$ with probability $1-\rho$.

For any trajectory $\Theta$, define the activity $K_t(\Theta)$ as the total number of (successful) hops in the time-interval $[0,t]$.  
 Within the steady state of the model, one has $\langle K_t/t \rangle_0 = 2L\rho(1-\rho)$, since the rate of attempted hops is $2N=2\rho L$ and the expected fraction of successful hops is equal to the probability ($1-\rho$) that the destination site is unoccupied.  We consider the distribution $\tilde P$ defined as in (\ref{eq:biastraj}), with $A_t=K_t$: from (\ref{equ:P-Theta}) one has
 $$
\tilde{P}_{\tmax}(\Theta,s) = \frac{1}{Z(s,\tmax)} \left[ \prod_{k=0}^{K-1}  W(\CC_k \rightarrow \CC_{k+1}) {\rm e}^{-s} \right] \cdot \exp \left[-\sum_{k=0}^K r(\CC_k) (t_{k+1} - t_{k }) \right] .
$$

\subsection{Modified dynamics}
\label{subsec:mod-dyn-ssep}

To improve computational efficiency when sampling from $\tilde{P}$, it is useful to adopt a simple modification to the dynamics, as described in Section~\ref{subsec:moddyn}.  
In this modification, all transition rates are rescaled by a factor ${\rm e}^{-s}$.

From (\ref{equ:P-Theta}), the resulting probability distribution is
$$
\hat{P}(\Theta) = \left[ \prod_{k=0}^{K-1}  W(\CC_k \rightarrow \CC_{k+1}) {\rm e}^{-s} \right] \cdot \exp \left[-\sum_{k=0}^K r(\CC_k) {\rm e}^{-s} (t_{k+1} - t_{k }) \right].
$$
Hence, from (\ref{eq:upsdef}),
\begin{equation} \Upsilon_{\tmax}(\Theta) = \exp\left[ -\sum_{k=0}^K r(\CC_k) (1-{\rm e}^{-s}) (t_{k+1} - t_{k }) \right] . 
\label{equ:def-ups-ssep}
\end{equation}

In the following, we use these modified dynamics and the weight factors $\Upsilon$ within our implementation of the cloning algorithm.
This modification to the dynamics is useful because the biasing field $s$ has the effect of suppressing all transitions, so as to reduce $K$.  Since the modified dynamics incorporates this (trivial) effect, the resulting trajectories are closer to the biased trajectories of interest than one would get by simulating the SSEP directly.

\subsection{Dynamical phase transition}

Our motivation for studying large deviations of $K_t$ in the SSEP is twofold. First, the model is simple enough for a precise numerical characterisation, and is a useful test of the numerical method.  Second, there is a dynamical phase transition that takes place in the model \cite{bodineau2007cumulants, bodineau2005distribution, lecomte2012inactive} which reveals interesting physical effects. 
To investigate the phase transition, it is useful to scale the bias $s$ by the square of the system size: we define
\begin{equation}
\lambda = sL^2
\label{equ:def-lambda}
\end{equation}
The phase transition takes place at a critical value of $\lambda$, and 
leads to numerical challenges that we use in later sections to test the cloning method. 

The phase transition appears if one fixes the density $\rho=N/L$ and takes the lattice size $L\to\infty$.  [We take this limit after the large-$t$ limit associated with the large deviation principle (\ref{eq:ldp}).]  For finite $L$, the rate function $\pi(a)$ and the cumulant generating function $\psi(s)$ in (\ref{equ:psi-z}) are both analytic functions, and there is no phase transition. However, on taking $L\to\infty$, a singular response to the field $s$ can be observed, just as conventional phase transitions can be observed on taking the thermodynamic limit.  Specifically, one considers $\psi_*(s)=\lim_{L\to\infty} [\psi(s)/L]$, which is analogous to a thermodynamic free energy density, whose derivatives show singular behaviour at phase transitions.

\begin{figure*} [t]
\includegraphics{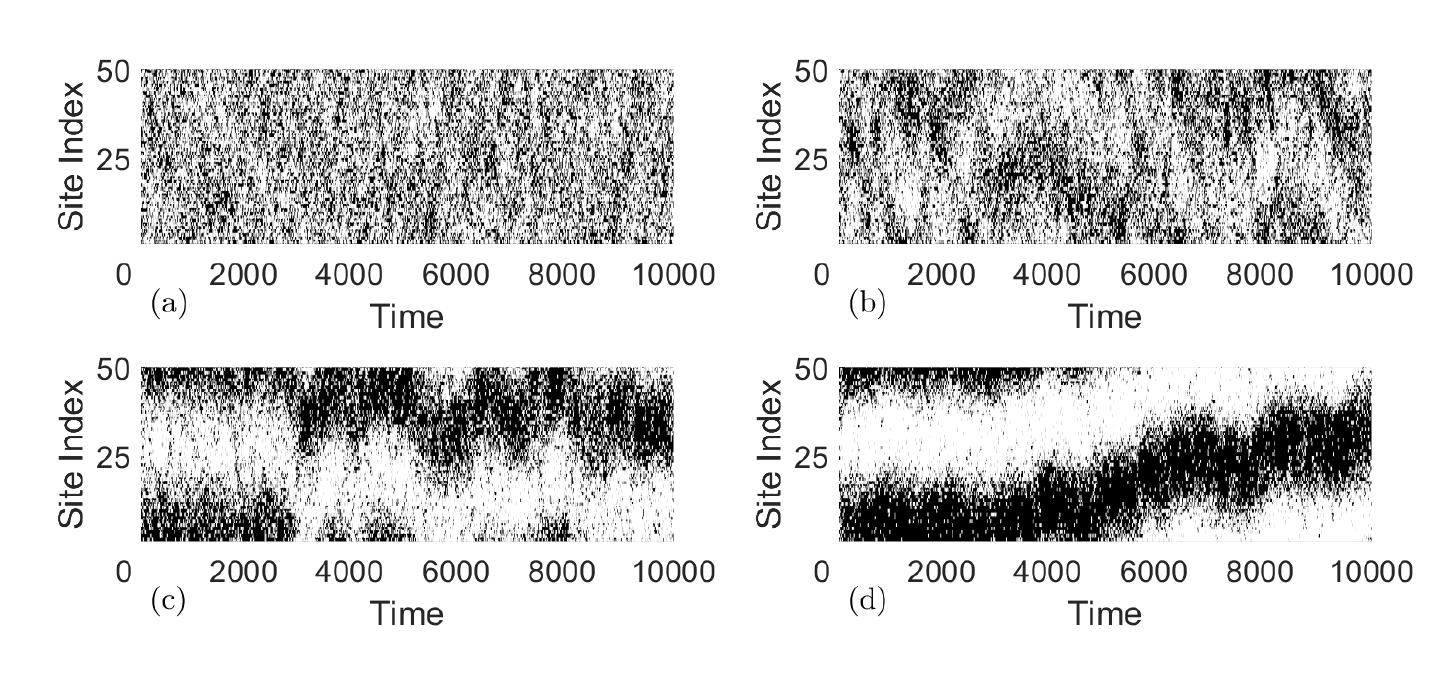}
\caption{Sample trajectories of SSEP with $L=50,\:N=L/2$ and $\tmax=10^4$.  (a)~$s=0$, the equilibrium state; (b)~$s=0.008$, showing evidence of transient clusters; (c)~$s=0.012$, in which a single large cluster has formed; (d)~$s=0.020$, with most of the particles in a well-defined single cluster.  The corresponding values of $\lambda$ are $0,20,30,50$ and the critical value of $\lambda$ is $\lambda_c = 2\pi^2\approx 19.7$.}
\label{fig:Traj}
\end{figure*}

The physical signature of this transition is shown in Figure \ref{fig:Traj} where there is a transition from a homogeneous state at $s=0$ (particles are distributed evenly throughout the system) to an inhomogeneous (``phase-separated'') state for $s>0$, in which case the particles are segregated into a dense and a dilute region.

To quantify the particle clustering that takes place for $s>0$, it is useful to consider the Fourier transform of the density field:
\begin{equation}
\label{eq:fourdens}
\delta \rho_n = \frac{1}{\sqrt{L}} \sum_{j=1}^N e^{- 2\pi {\rm i} n X_j/L}
\end{equation}
where $X_j$ is the index of the site occupied by particle $j$, and $n=0,1,\dots L-1$.  We focus on the wave vector that corresponds to the longest wavelength fluctuations, that is $n=1$.

\begin{figure*} [t]
\centering
\includegraphics[width=8cm,height=6cm]{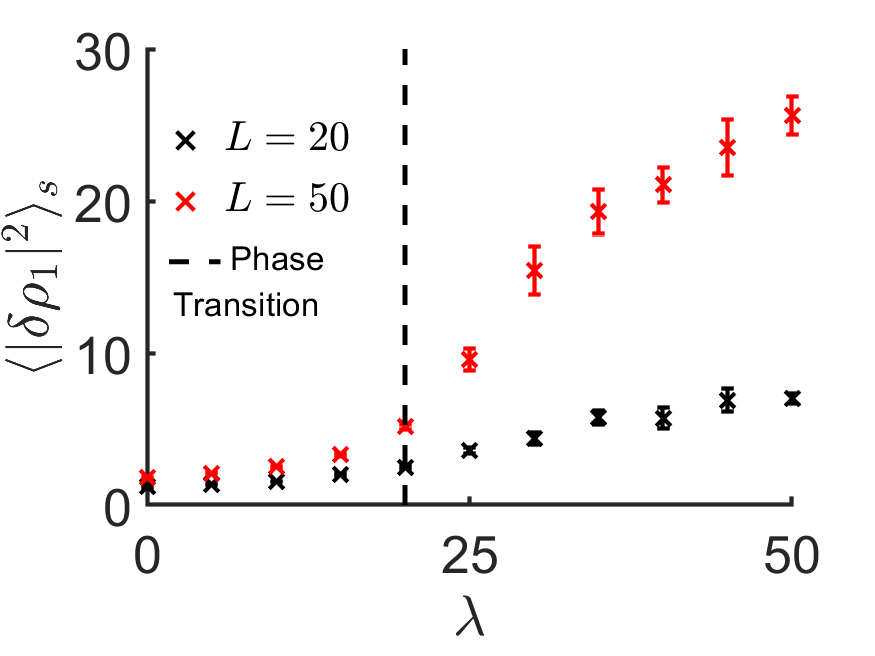}
\caption{The mean square value of the first Fourier component of the density, $\langle |\delta \rho_1|^2 \rangle_s$, measured at $t=\tmax/2$ with $\tmax=10^4,\Nc=10^5$, for various $L,\lambda$.  The phase transition occurs at $\lambda_c=2\pi^2$: for $\lambda>\lambda_c$ one expects the system to become inhomogeneous, so that $\langle |\delta \rho_1|^2 \rangle_s \propto L$, consistent with the data.}
\label{fig:dpqGraph}
\end{figure*}

Figure \ref{fig:dpqGraph} shows that the magnitude of this Fourier component grows as the system becomes inhomogeneous.

\subsection{Scaling at the dynamical phase transition}
\begin{figure*} [t!]
\includegraphics{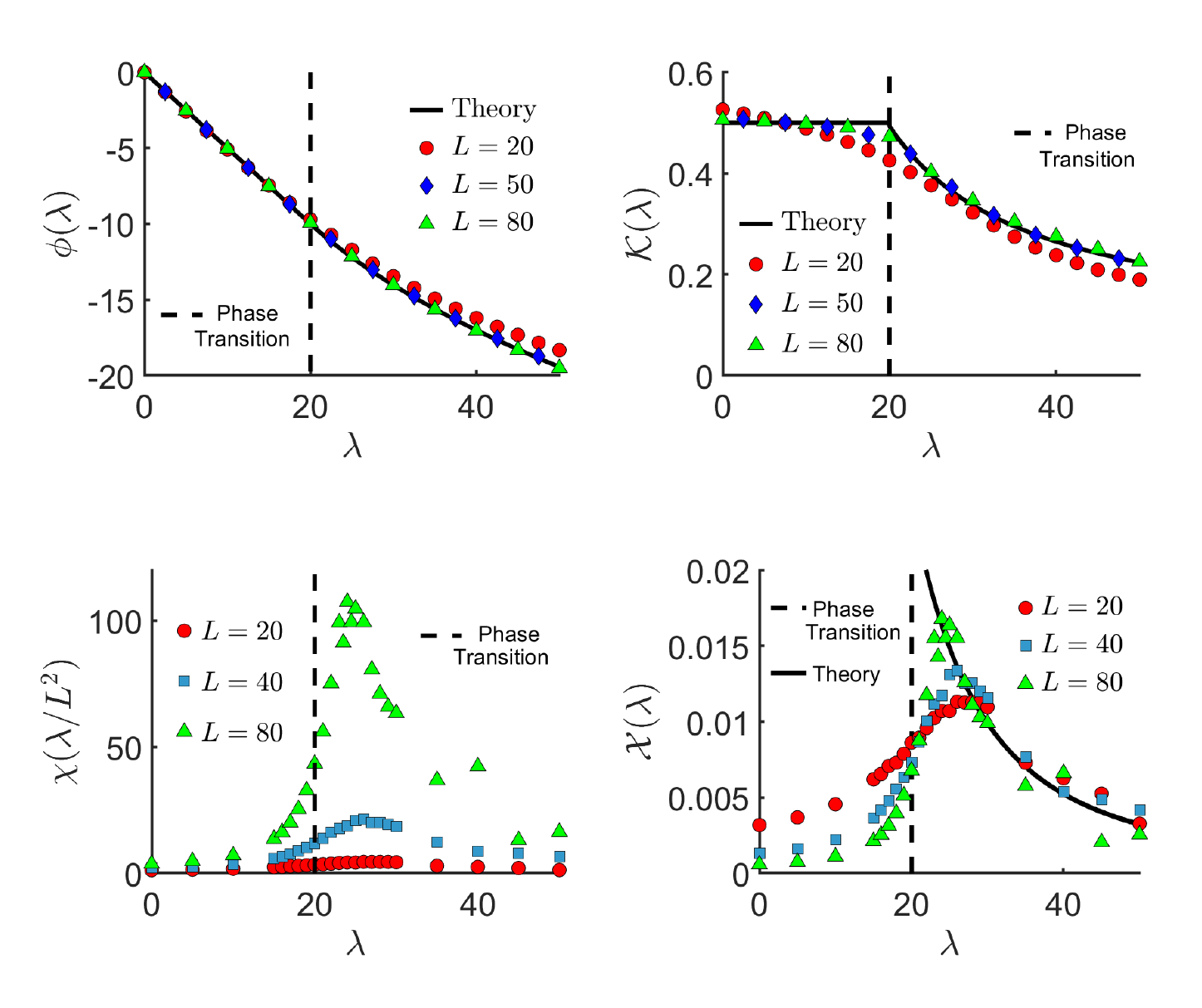}
\caption{Results illustrating the dynamical phase transition in the SSEP. The number of clones is $\Nc=10^5$ for $L=20,40,50$ and $\Nc=10^6$ for $L=80$. The cloning interval is $\Delta t = 10$  and all results are averaged across 10 independent computations. The vertical dashed line shows the position of the phase transition ($\lambda_c=2\pi^2$).  The theoretical predictions are obtained by using finite differences to take derivatives of $\phi(\lambda)$, which is calculated as described in~\cite{lecomte2012inactive}. For $\lambda<\lambda_c$ one has $\mathcal{X}\to0$ as $L\to\infty$: the leading behaviour for large $L$ is ${\cal X}=O(1/L)$, and was computed in~\cite{appert08}.  Our numerical results are consistent with that prediction (data not shown).}
\label{fig:AlgRes}
\end{figure*}

To investigate this phase transition in more detail, it is useful to ``zoom in'' on the crossover from the homogeneous to the inhomogeneous case.  To achieve this, let $s=\lambda/L^2$ as in (\ref{equ:def-lambda}) and consider the limit of large $L$ at fixed $\lambda$. 
This is analogous to finite-size scaling in equilibrium systems.  
We focus in this work on the crossover function
\begin{equation}
\KK_L(\lambda) = k({\lambda}/{L^2}) , \qquad \hbox{where}\qquad k(s) = L^{-1}\lim_{t\to\infty}\left< K_t/t \right>_s = -\psi'(s)/L .
\label{equ:def-calK}
\end{equation}
As $L\to\infty$, the function $\KK_L$ converges to a limiting form $\KK_*$, which can be computed using macroscopic fluctuation theory~\cite{lecomte2012inactive}.  This function has a singularity at $\lambda=\lambda_c=2\pi^2$.  For $\lambda<\lambda_c$, one has a constant value $\KK_*=2\rho(1-\rho)$.  For $\lambda>\lambda_c$, the function $\KK_*$ decreases with $\lambda$, converging to zero as $\lambda\to\infty$.
This behaviour is illustrated in Figure~\ref{fig:AlgRes}, which also shows the rescaled ``free energy''
$$
\phi_L(\lambda) = L \psi(\lambda/L^2)
$$
To show the singularity that appears at the phase transition, we also show two measures of susceptibility
\begin{equation}
\chi_L(s) = L^{-1} \psi''(s), \qquad \XX_L(\lambda) = -\KK_L'(\lambda) = L^{-2} \chi_L(\lambda/L^2).
\label{equ:def-chi}
\end{equation}
The bare susceptibility $\chi_L$ corresponds to the scaled variance of $K_t$ (recall (\ref{equ:psi-var})), and $\lim_{L\to\infty}\chi_L(s)$ should have a finite value in a system that is away from any phase transition.  Figure~\ref{fig:AlgRes} shows a divergence in $\chi$, consistent with the existence of a phase transition.  On the other hand, the function $\XX_L(\lambda)$ is predicted by MFT to have a finite limit as $L\to\infty$.  Our numerics are consistent with this prediction but they show that measuring this limiting function requires large system sizes.

Note that the scenario illustrated  in Figure~\ref{fig:AlgRes} is different from classical finite-size scaling and from other first-order dynamical phase transitions~\cite{nemoto2017}.  Comparing with the classical case, note that we take $\tmax\to\infty$ and then later $L\to\infty$.  As discussed in Ref.~\cite{nemoto2017}, this is equivalent to thermodynamic finite-size scaling in a cylindrical geometry, with the length of the cylinder being much longer than its perimeter.  In that case, one possibility is that the susceptibility $\chi_L$ grows exponentially in the system size, due to the a distribution of domains along the cylinder, with ``domain walls'' perpendicular to the long axis of the cylinder.
However, the results presented are very different from that case: one reason is that the coexisting phases at the transition have different densities, but the number of particles in the SSEP is equal at every time.  As a result, the ``domain walls'' between dense and dilute regions in Figure~\ref{fig:Traj} are constrained to lie parallel to the time axis.  In general, $\chi_L$ can be interpreted as a ``correlation volume'' in space-time.   From (\ref{equ:def-chi}), and noting that $\mathcal{X}(\lambda)$ is finite (or zero) for all $\lambda$~\cite{lecomte2012inactive}, one sees that $\chi_L$ diverges as $L^2$ in the vicinity of the transition.  This factor arises from the characteristic time scale proportional to $L^2$ that is associated with density fluctuations on length scale $L$.  Note also that while $k(s)$ exhibits a jump at $s=0$, corresponding to a first-order phase transition, one may also consider the behaviour of the system as a function of $\lambda$, in which case $\mathcal{K}(\lambda)$ is continuous at the transition but has a discontinuous first derivative: when viewed on this scale, the transition has some features of a continuous phase transition~\cite{espigares17,baek17}.

\section{Implementation of algorithm and cloning stage}
\label{sec:Algorithm}

We outlined the cloning algorithm in Section~\ref{subsec:alg}.  Here we provide some extra detail on its application to the SSEP.
The modified SSEP dynamics are implemented using the Bortz-Kalos-Lebowitz (continuous time Monte Carlo) algorithm~\cite{bortz75}.  All possible particle hops have the same rate ${\rm e}^{-s}$.  These dynamics take place over the time interval $[t_{\beta-1},t_{\beta}]$.  During this time period, the factor $\Upsilon^\beta(\Theta_i)$ is calculated: based on (\ref{equ:def-ups-ssep},\ref{eq:upsbdef}) one may write
$$\Upsilon^{\beta}(\Theta_i) = \exp \left( -\int_{t_{\beta-1}}^{t_{\beta}} (1-{\rm e}^{-s})r(\CC_t^i) dt \right)$$
where $\CC_t^i$ is the configuration of clone $i$ at time $t$ and $r(\CC)$ is the escape rate for configuration $\CC$ under the original (unmodified) dynamics.  The factors $\Upsilon$ appear in the estimate (\ref{equ:psihat}) for $\psi(s)$.

\subsection{Copying and deletion of clones: [eq] and [iid] methods.}
\label{sec:alpha}

The next step is to clone and delete systems, according to their values of $\Upsilon$, so as to produce a new population.  
There is some freedom as to how this is implemented within the algorithm. What we require is that for a large population the average number of descendants of clone $i$ approaches $\Upsilon^\beta(\Theta_i)/\Upsilon^\beta_{\rm T}$ with $\Upsilon_{\rm T}=\sum_{i=1}^{\Nc} \Upsilon^\beta(\Theta_i)$.  We consider two methods for selecting the clones that will form the new population, both of which are consistent with this requirement. 

\begin{figure}
\includegraphics{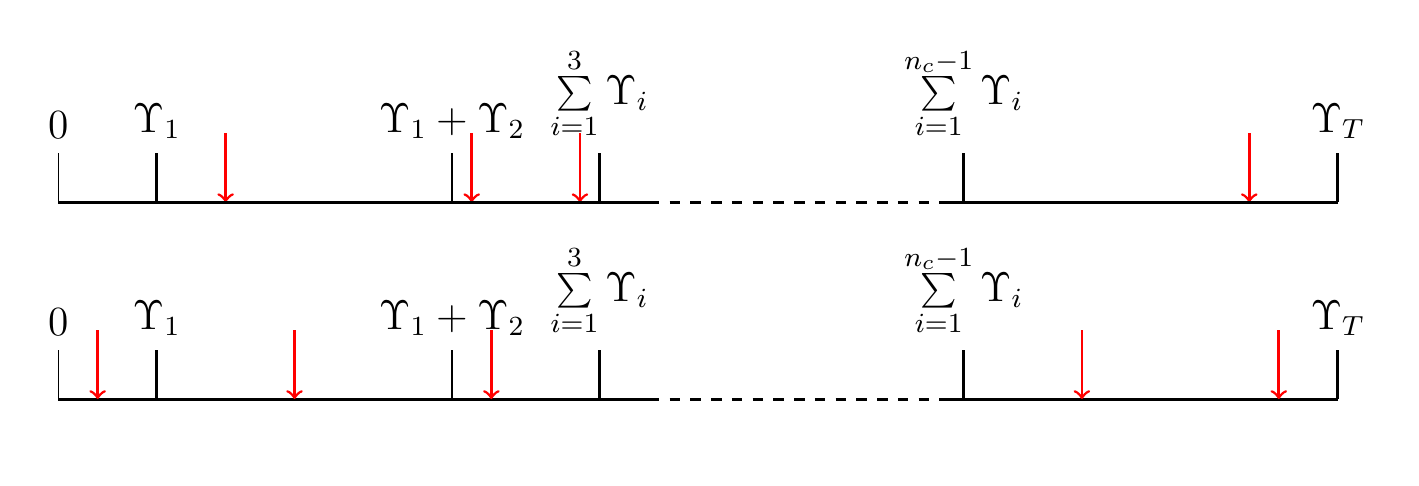}
\caption{Example of the number line associated with the cloning factors $\Upsilon^\beta(\Theta_i)$ used in the clone selection methods, as described in the main text.  We take $\Nc=1000$. The red arrows indicate some of the points $\alpha_j$ that determine which clone is selected for inclusion in the new population.  The $\alpha$ may be chosen independently ([iid] method, top line) or equally spaced ([eq] method, lower line).}
\label{fig:Key}
\end{figure}

The first clone selection method is denoted by [eq]: the reason for this will be explained below.  In this method, for each clone $j$ of the new population (with $1\leq j\leq \Nc$), we define $\alpha_j=(j+d-1)\Upsilon_{\rm T}/\Nc$, where $d$ is a random number in $(0,1)$ which is equal for each clone.  Thus, the $\alpha_j$ are equally spaced on $[0,\Upsilon_{\rm T}]$, with $\alpha_1=(d\Upsilon_{\rm T}/\Nc)$ and $\alpha_{\Nc}=\Upsilon_{\rm T}-[(1-d)\Upsilon_{\rm T}/\Nc]$.  Then the state of clone $j$ of the new population is pulled from clone $k$ of the old population, where $k$ satisfies 
\begin{equation}
\sum_{i=1}^k \Upsilon^\beta(\Theta_i) \leq \alpha_j < \sum_{i=1}^{k+1} \Upsilon^\beta(\Theta_i).
\label{equ:alpha-Ups}
\end{equation}
This clone selection method is equivalent to constructing a line segment of total length $\Upsilon_{\rm T}$ that is composed of contributions from each clone of the old population, with clone $k$ contributing a length $\Upsilon^\beta(\Theta_k)$.  Then clone $j$ of the new population is selected by selecting the interval of the original line that contains the point $\alpha_j$ as in Figure \ref{fig:Key}.  On average, the number of copies of clone $k$ in the new population is then $\Upsilon^\beta(\Theta_k)/\Upsilon_{\rm T}$, as required.  The label [eq] is used because the $\alpha_j$ are equally spaced on the interval $[0,\Upsilon^\beta_{\rm T}]$.

The second clone selection method that we use is denoted by [iid].  In this case, the $\alpha_j$ are identically and independently distributed (uniformly) on $[0,\Upsilon^\beta_{\rm T}]$, hence the label [iid].
However, this choice is less efficient, as we discuss below (Section~\ref{sec:Interval}). As noted in Sec.~\ref{subsec:alg}, the two sub-steps, of independent dynamical evolution followed by cloning, are each repeated $M$ times, so that the total simulation time for each clone is $\tmax$.  The final step is a cloning step.  

\subsection{Test case}
\label{sec:test-case}

As a stringent test of algorithmic performance, we focus on the observable $\mathcal K_L(\lambda)$ defined in (\ref{equ:def-calK}).  We consider values of $\lambda$ in the vicinity of the phase transition.

One sees from Figure~\ref{fig:AlgRes} that this function has a feature at $\lambda=2\pi^2$ that depends strongly on system size.  In the following,  we will test the ability of the algorithm to capture the finite-size scaling of this feature.
Our numerical estimator for $\mathcal{K}_L$ is 
\begin{equation}
\label{equ:khat}
\hat{k}_L(\lambda)=\frac{1}{L \tmax} \sum_{i=1}^{\Nc} \sum_{\beta=1}^M K^{\beta}(\hat{\Theta}_i).
\end{equation}
see also (\ref{equ:psihat}): we again emphasise that in writing $K^\beta$ then $\beta$ is an index (not an exponent).  Note, that computational evaluation of $\hat{k}$ requires that the trajectories $\hat\Theta$ can be obtained by extrapolation backwards in time: this is easily achieved by including with its clone its accumulated value of $\sum_\beta K^\beta(\hat\Theta_i)$, which is copied along with the system's configuration when the clone state is copied, during the cloning stage of the algorithm (see~\cite{tailleur09aip}).  We use the method to measure $\cal K$ since this is a more revealing measure of algorithmic errors than $\psi(s)$ or $\phi(\lambda)$, as may be seen already in Figure~\ref{fig:AlgRes}.  An alternative estimator of $\cal K$ can be obtained by using finite differences to estimate the derivative $\phi'(\lambda)$: we prefer the direct measurement (\ref{equ:khat}), which avoids uncertainties arising from the finite differencing, although it does require an accurate sampling of the distribution $\pave(K^\beta)$ which may be challenging in practice: see Sec.~\ref{subsec:ncConv} below.    

In the following, we particularly focus on systematic errors: we average $\hat{k}$ over $R$ independent computations (indexed by $r=1,2,\dots R$) and we denote the average of the estimator by
$$\overline{k}_L(\lambda)=\frac{1}{R} \sum_{r=1}^R \hat{k}^r_L(\lambda),$$ 
and its variance by 
$$[\Delta k_L(\lambda)^2]=\frac{1}{R} \sum_{r=1}^R \left[\hat{k}^r_L(\lambda)^2-\overline{k}_L(\lambda)\right]^2.$$
Consequently, its standard deviation is $\sigma(\hat{k})=[\Delta k_L(\lambda)^2]^{1/2}$.  
For large $R$, the systematic error of the method can be obtained as the difference between $\overline{k}_L(\lambda)$ and $\mathcal{K}_L(\lambda)$, and the size of the random error is determined by the variance $[\Delta k_L(\lambda)^2]$.

We emphasise that $\hat{k}_L(\lambda)$ is a simple estimate of ${\cal K}_L(\lambda)$ -- other estimates might be obtained by running the algorithm with different parameters (for example a range of $\Nc$ or a range of $\lambda$), and combining the data in order to extrapolate or interpolate an improved estimate.  See also Sec~\ref{sec:Conv} below.  However, when analysing the errors of the algorithm we concentrate on the direct estimate $\hat{k}_L$, since this can be analysed in a simple and precise way (for example, the use of a direct estimate means that there are no correlations of the errors between different state points in the following Figures).

\begin{figure*} [t]
\includegraphics[width=15.cm]{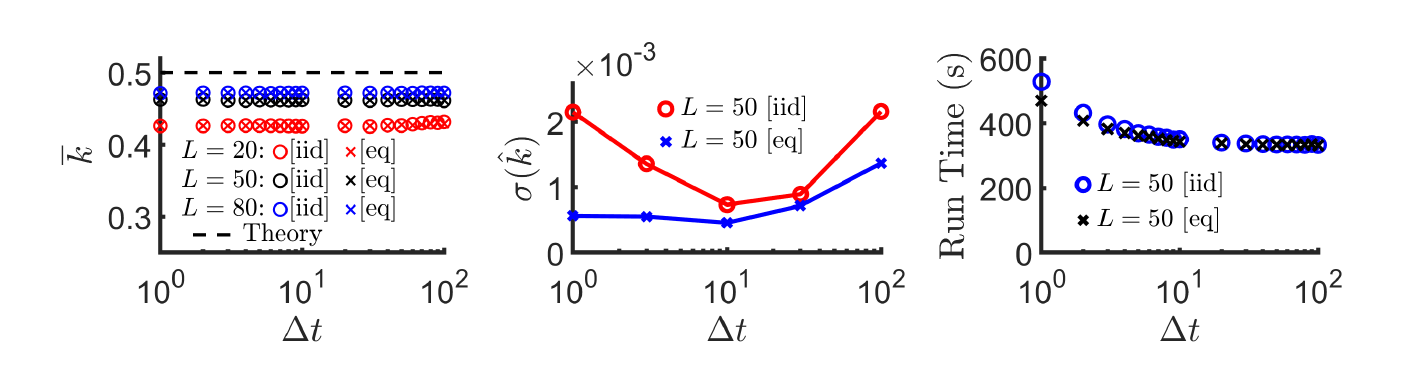}
\caption{Dependence of results on the cloning interval $\tau$ and the clone selection mechanism ([eq] or [iid]) at the representative state point $\lambda=28$ (see main text).  (a)~Analysis of the two clone selection mechanisms, showing that if the simulation is converged, the estimator $\overline{k}$ is (almost) independent of $\Delta t$, and of the clone selection method (as required). (b) The statistical errors are heavily dependent on the clone selection method and vary with $\Delta t$. (c) The computational time is almost independent of the clone selection method.  It decreases with $\Delta t$ because the amount of communication between nodes decreases, but this has a cost in terms of accuracy (see panel (b)).
}
\label{fig:Tau}
\end{figure*}

\subsection{Effect of cloning method and choice of $\Delta t$}
\label{sec:Interval}

Given a large enough number of clones, the results of the algorithm are independent of the parameter $\Delta t$, and they are also independent of whether the $\alpha_j$ are chosen to be equally spaced [eq] or uniformly at random [iid]. In this section, we show the effects that these choices have on the results obtained with finite $\Nc$. As a representative state point (unless otherwise stated) we focus on $\lambda=28$ and $\tmax=10^4$ with $\Delta t=10$, but the results are qualitatively similar for other parameters too. (Note that in this section we show data for both [eq] and [iid] clone selection methods, to illustrate that the performance of the [eq] method is better.  All other sections use the [eq] method, unless stated otherwise.)

Results are shown in Figure~\ref{fig:Tau}.  Panel (a) shows that on using sufficiently many clones and averaging over many computational realisations, the results are independent of $\Delta t$, as expected.  Panel (b) shows how the variance of the estimate of $\KK_L$ depends on $\Delta t$: the smaller is this variance, the less computational realisations are required to get accurate results, and hence the algorithm is more efficient.  For the [eq] method, the variance is always smaller than for the [iid] method.  Moreover, for [eq] method, the variance is monotonically increasing as a function of $\Delta t$.  This implies that using a small value of $\Delta t$ is desirable, from the point of view of accuracy.  In practice, computational requirements mean that $\Delta t$ should not be too small (Figure~\ref{fig:Tau}(c)), since the cloning stage of the algorithm incurs an overhead.  For the [eq] method, this leads to a simple trade-off between accuracy (better for small $\Delta t$) and computational efficiency (better for large $\Delta t$).  For the [iid] method, this trade-off is more complicated since accuracy may also be reduced by choosing smaller $\Delta t$.

To see why the [eq] method is more efficient, it is useful to consider the operation of the algorithm with $s=0$.  In this case, the most efficient algorithm is clearly to simulate each copy independently, so as to obtain $\Nc$ independent samples.  Choosing equally-spaced $\alpha_j$  ensures that this does indeed happen (all the $\Upsilon^\beta(\Theta_k)=1$ so one $\alpha_j$ lies in each interval).  However, choosing the $\alpha_j$ independently means that some copies of the system are copied several times and some are deleted, because of the randomness inherent in the choice of  $\alpha_j$.  The deletion of clones reduces the number of independent samples in the system and tends to increase the errors.  The effect is the same for non-zero $s$ and the problem also occurs in the continuous-time cloning algorithm as described in~\cite{lecomte2007numerical}.  Hence, our conclusion is that choosing equally-spaced $\alpha_j$ as described in Sec.~\ref{sec:Algorithm} is the more accurate of our two methods for selecting clones.

More generally, it is desirable -- while always maintaining algorithmic accuracy -- to minimise the number of clones that are deleted, since each deletion results in an (irreversible) loss of information, reducing the number of independent samples from $\tilde{P}$.

\section{Dependence of the results on the parameters of the cloning algorithm}
\label{sec:Conv}

We have emphasised that the cloning algorithm gives accurate results only in the limit $\Nc\to\infty$.  Characterising large deviations of the activity also requires that $\tmax\to\infty$.  Our ability to probe these limits depends on the available computational resource: it is therefore essential to characterise and understand the dependence of the algorithm's results on $\Nc$ and $\tmax$, in order to assess whether the algorithm gives reliable results. For finite $\Nc$, a suitable choice of $\Delta t$ is also essential to obtain accurate results.  This section describes the dependence of the results on $\Nc$ and $\tmax$.

We will find that the convergence of the algorithm depends significantly on the value of $\lambda$, particularly whether the system is in the homogeneous regime $\lambda<\lambda_c$ or the phase-separated regime $\lambda>\lambda_c$.  The convergence also depends on the system size $L$, with large values of $\Nc,\tmax$ being required when $L$ is larger.  When assessing the accuracy of our results, we focus on systematic errors, and we aim to achieve a relative error of less than $2\%$ on our estimates of ${\cal K}(\lambda)$.

\subsection{Convergence with respect to the time $\tmax$: effects of long time scales}

\begin{figure*}[t]
\includegraphics{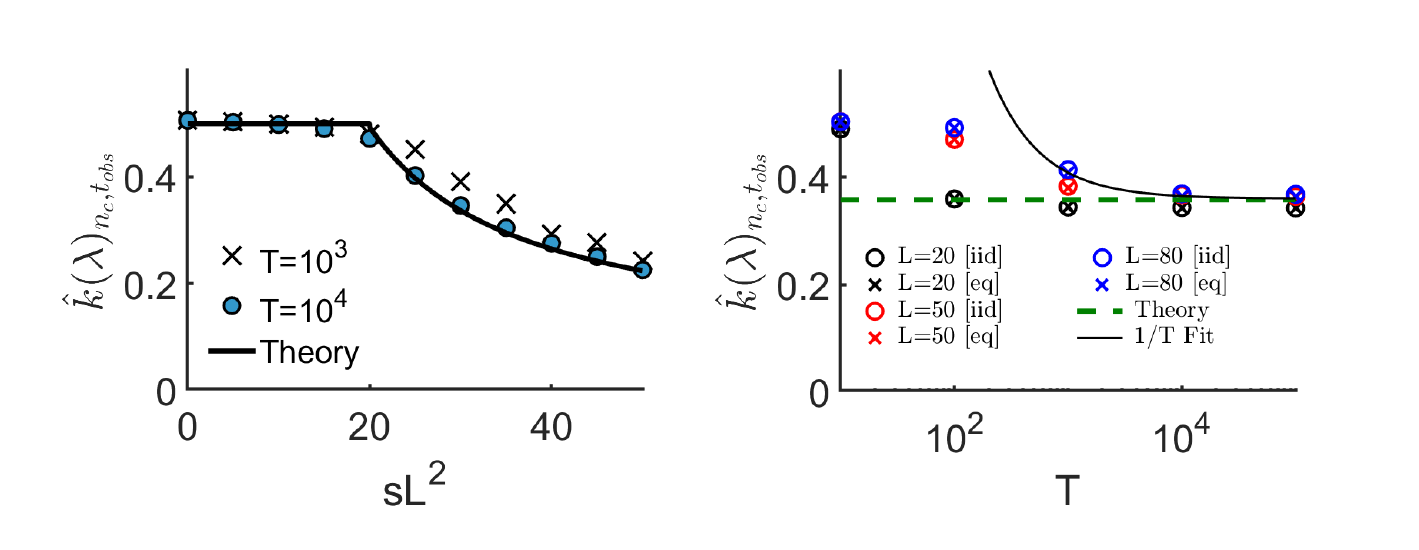}
\caption{Estimates $\overline{k}(\lambda)$ of the activity $\mathcal{K}_L(\lambda)$ as a function of $\tmax$ for various system sizes $L$.  (a)~$L=50$ for various $\lambda,\tmax$.  (b)~$\lambda=28$ for various $\tmax,L$. 
The ``theory'' line is $\mathcal{K}_* = \lim_{L\to\infty} \mathcal{K}_L(\lambda)$.
We took $R=10$, with $\Nc=10^5$ for $L=20,50$ and $\Nc=10^6$ for $L=80$ (these $\Nc$ are large enough that results depend very weakly on $\Nc$). 
The fit uses equation (\ref{equ:fit-finite-tmax}) applied to the data for $L=80$ [eq], using only the points at $\tmax=10^3,10^4$.  This fit captures the scaling for large $\tmax$.  See also Table~\ref{table:efficiency-T}.
}
\label{fig:T}
\end{figure*}

To analyse the dependence of the method on the choice of $\tmax$, we consider the dependence of $\overline{k}(\lambda)$ on the observation time $\tmax$.  In this section, the number of clones used is sufficiently large that the results depend very weakly on $\Nc$: see Section~\ref{subsec:ncConv}.  Results are shown in Figure~\ref{fig:T}.  For $\lambda<\lambda_c$, Figure~\ref{fig:T}(a) shows that results depend weakly on $\tmax$ as long as $\tmax\gtrsim 10^3$.  On the other hand, for $\lambda>\lambda_c$ there is a systematic dependence on $\tmax$ even for $\tmax>10^3$.  This message is confirmed by Figure~\ref{fig:T}(b) which shows that systematic errors from finite $\tmax$ are larger for larger systems, but $\tmax=10^4$ seems to be large enough to achieve convergence even for $L=80$.
To verify this effect, we show (for $L=80$) a fit to the asymptotic prediction~\cite{nemoto17finite,hidalgo17finite} 
\begin{equation}
\bar{k}(\lambda)_{n_c,\tmax} = k_\infty ( 1 + A/\tmax ) + O(\tmax^{-2}) 
\label{equ:fit-finite-tmax}
\end{equation}
The fit (with parameters $k_\infty,A$) is performed using the data points at $\tmax=10^3,10^4$ and the resulting fit is consistent (up to our $2\%$ tolerance) with the measured data point at $\tmax=10^5$.  Thus, the results are consistent with the theory of~\cite{nemoto17finite,hidalgo17finite}, and this fitting also provides accurate estimates of $k_\infty$, provided one performs the fit using data points that are within the asymptotic regime.  We show this fit for $L=80$ and the [eq] clone selection method, the results for other cases are similar, as summarised in Table~\ref{table:efficiency-T}.  Note however that data for small $\tmax$ ($\lesssim100$) are not at all consistent with the asymptotic prediction (\ref{equ:fit-finite-tmax}): one should remember that if data for large $\tmax$ are not available, it may be difficult to assess which data are representative of the asymptotic regime and which should be excluded from the fit.

\begin{table}
\centering
\begin{tabular}{ c |  c | c | c | c | c | c |}
     & \multicolumn{2}{c} {L=20} \vline & \multicolumn{2}{c} {L=50} \vline & \multicolumn{2}{c} {L=80}\\ 
     \cline{2-7}
     & [iid] & [eq] & [iid] & [eq] & [id] & [eq]\\
     \hline
     Fit parameter $A$ (exc. $\tmax=10^5$) & 1.0 & 2.6 & 17.9 & 18.6 & 50.1 & 48.4 \\
     Fit parameter  $k_\infty$ (exc. $\tmax=10^5$) & 0.3437 & 0.3434 & 0.3657 & 0.3620 & 0.3645 & 0.3605\\
     Fit parameter $k_\infty$ (incl. $\tmax=10^5$) &  0.3433 & 0.3432 & 0.3649 & 0.3628 & 0.3666 & 0.3629\\
     Difference in $k_\infty$ & $0.1\%$ & $<0.1\%$ & $0.2\%$ & $0.2\%$ & $0.6\%$ & $0.7\%$ \\ 
\end{tabular}
\caption{Results of fitting the results in Fig.~\ref{fig:T}(b) to Equ~(\ref{equ:fit-finite-tmax}), using data for $\tmax\geq10^3$.  The first two rows are results of fitting just two points $\tmax=10^3,10^4$.  The third row shows the estimate of $k_\infty$ when the final point at $\tmax=10^5$ is included in the fit.  The asymptotic prediction for $k_\infty$ is robust as more data are added, indicating that the data are consistent with the fit.}
\label{table:efficiency-T}
\end{table}

To understand the physical origin of the errors that arise from finite $\tmax$, recall (\ref{eq:TTIave}-\ref{eq:Opend},\ref{equ:khat}) and also the discussion of Section \ref{subsec:measave}.  The trajectories sampled by the algorithm have transient regimes at initial and final times, whose typical time scale is denoted by $\tau$.  In terms of (\ref{equ:khat}), this means that for values of $\beta$ that are outside the transient regimes, the variables $K^\beta$ are all distributed according to the distribution $\pave$: in this case the average of $K^\beta$ is independent of $\beta$ and (assuming that $\Nc$ is large enough) this average is equal to $\Delta tL\mathcal{K}_L(\lambda)$.  If the terms that are distant from the temporal boundaries dominate the sum in (\ref{equ:khat}) then $\overline{k}(\lambda)= \mathcal{K}_L(\lambda)$, as required. However, when $\beta$ is close to $1$ or $M$, the average value of $K^\beta$ depends on $\beta$ and is not equal to $\Delta tL\mathcal{K}_L(\lambda)$.  The inclusion of these terms in (\ref{equ:khat}) means that $\overline{k}(\lambda)\neq \mathcal{K}_L(\lambda)$ in general: there are corrections of the order of $\tau/\tmax$ coming from the transient regimes. 

\begin{figure*} [t!]
\includegraphics{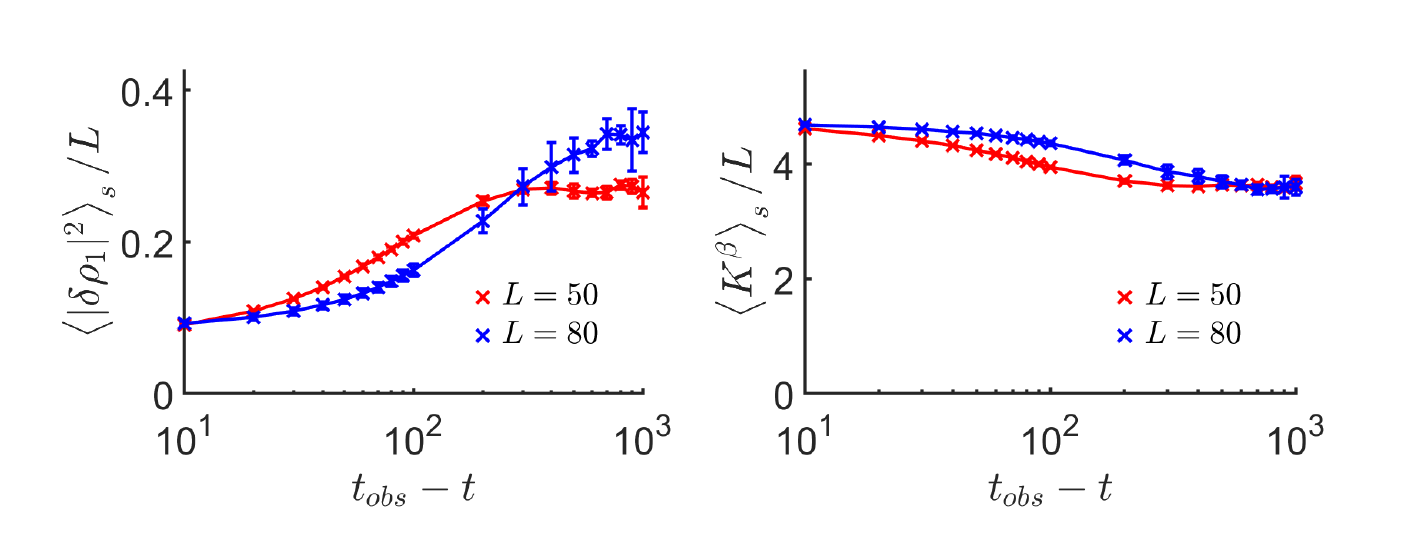}
\caption{Time-dependent averages of the form $\langle F_t \rangle_s$ with $F$ being either $|\delta \rho_1|^2$ (the squared modulus of the first Fourier component of the density) or $K^\beta$ (which is the number of particle hops between times $t=t_\beta$ and $t_\beta+\Delta t$.)  Parameters are $\lambda=28$, $\Nc=10^6$, $\tmax=10^4$, $\Delta t=10$.}
\label{fig:obsAve}
\end{figure*}

In fact, one can reduce these errors by excluding some of the transient terms from the sum.
Figure~\ref{fig:obsAve} shows the transient regime that occurs for $(\tmax-t)\lesssim\tau$, for two different observables.  Two features are important here: the time scale associated with the transient regime, and the difference between $\langle F_{\tmax} \rangle_s$ and $F_\infty$ (which is the value when $\tmax-t\gg\tau$, recall (\ref{eq:TTIave})).  For both observables, the order of magnitude of the transient time scale is $\tau\simeq100$ for $L=50$ and $\tau\simeq300$ for $L=80$.  Obtaining an accurate estimate of $\mathcal{K}$ from (\ref{equ:khat}) requires that $\tmax\gg\tau$, in order that the sum is dominated by terms that are outside the transient regime.  Alternatively one can estimate $\mathcal{K}$ from the plateau value of $\langle K^\beta\rangle_s/L$ at large $\tmax -t$ (the value should be normalised by a factor of $\Delta t=10$, for consistency with (\ref{equ:khat}).) 
This amounts to excluding transient terms from the definition of $\hat{k}$, and does indeed give accurate results, at the expense of some post-processing. (We  also note that the data in Fig.~\ref{fig:obsAve} are averaged over several independent runs of the algorithm, so identifying the plateau from the output of a single run of the algorithm may be non-trivial in practice.)  In the following we continue to analyse the simplest estimator $\hat{k}$  but we note that excluding transient terms from the sum in (\ref{equ:khat}) may well be a useful strategy for future applications of this algorithm.

Turning to the range of values of $\langle F_t\rangle_s$ as $\tmax-t$ is varied in Fig.~\ref{fig:obsAve}, one sees that $|\rho_1|^2$ changes by more than a factor of 2, while the fractional changes in $K^\beta$ are much smaller.  This indicates that it is the long-wavelength density modes that respond most strongly to the field $\lambda$, consistent with Figure~\ref{fig:dpqGraph}.  This is as expected, because these slowly-relaxing long-wavelength modes are the origin of the dynamical phase transition~\cite{thompson2015dynamical}.


\begin{figure*} [t]
\includegraphics{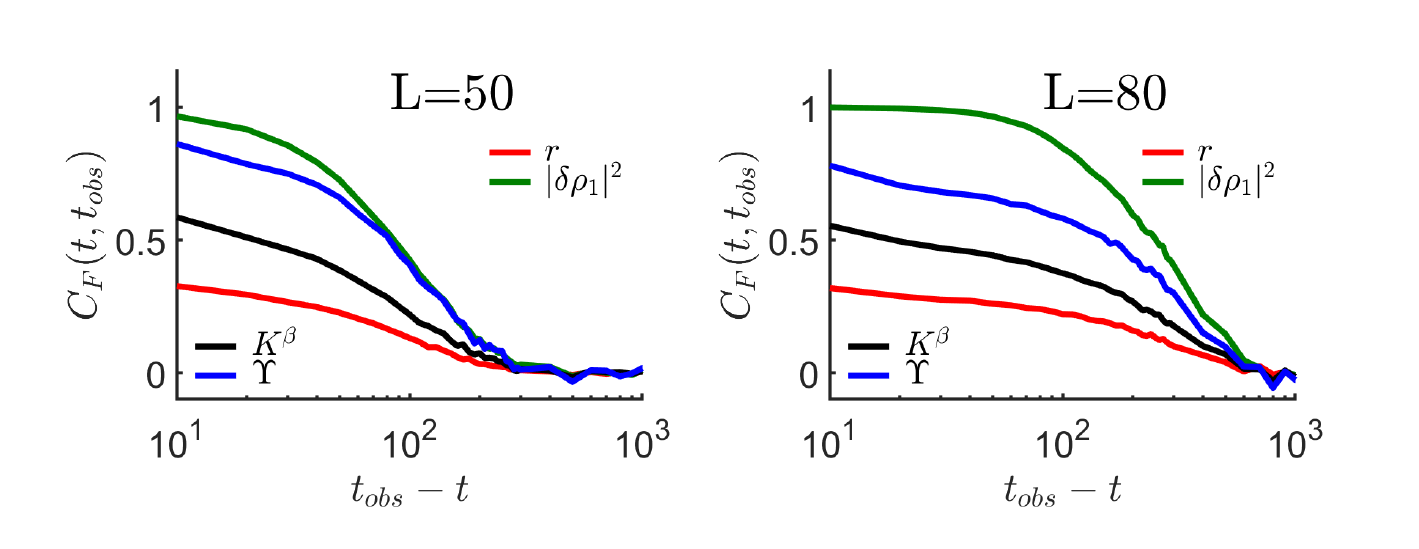}
\caption{Autocorrelation function $C_F$, as a function of the general time $t$ and the final time $\tmax$, for systems of $L=50$ (left), 80 (right). The normalisations $c(K_{\beta},t')$ are 946 ($L=50$) and 1450 ($L=80$). Data are shown for the observable $F$ being the escape rate $r$, the first Fourier component of density $|\delta p_1|^2$, the activity per cloning interval $K_{\beta}$ and the cloning factor $\Upsilon$. Parameters are $\tmax=10^4,~\Nc=10^6,~\lambda=28$, $\Delta t=10$. }
\label{fig:AutoCorr}
\end{figure*}

To understand these dynamical fluctuations in more detail, we also
consider the autocorrelation function for observable $F$, defined as
\begin{equation}
\label{eq:AutoCorr}
C_F(t,t') = \frac{\langle F_{t'} F_{t} \rangle_s - \langle F_{t'} \rangle_s \langle F_t \rangle_s}{c_F(t')}
\end{equation}
where $c_F(t') = \langle F_{t'} F_{t'} \rangle_s - \langle F_{t'} \rangle_s \langle F_{t'} \rangle_s$ is a normalisation factor that ensures that $C_F(t,t)=1$.  Since the transient time $\tau$ is comparable with the inverse of the spectral gap of the stochastic process~\cite{jack2010large}, one expects $C_F(t,t')$ to be small if $t-t'\gg\tau$.  Figure~\ref{fig:AutoCorr} shows results for several different observables, always with $t'=\tmax$ (this is the case for which good statistics are most easily obtained).  For the Fourier component $F=|\delta\rho_1|^2$, one sees that $C_F(t,\tmax)$ remains close to 1  until $\tmax-t\approx \tau$, after which point the density fluctuations at the two times decorrelate and the correlation decays.  On the other hand, for other observables such as the activity $K^\beta$, the correlation $C(t,\tmax)$ is significantly less than $1$ already for $\tmax-t=10$, indicating that the activity fluctuations have a ``fast component'' whose correlations decay quickly, as well as a slow component that is (presumably) correlated with the slow decay of large density fluctuations. 

We also note that the activity correlations play a special role in the theory: one has  $\chi(s) \sim L^{-1} c_F(\tmax) \int_{0}^{\tmax} C_{K^\beta}(t,\tmax) \mathrm{d}t$~\cite{garrahan2009first}, so $\chi$ can be large if either the prefactor $c_F$ is large, or if the function $C_{K^\beta}$ decays slowly to zero (so that the time integral becomes large).  The data in Figure~\ref{fig:AutoCorr} (and its caption) show that the prefactor $c_F$ scales roughly as $L$ while the time $\tau$ is expected to scale as $L^2$ (the slowest time scale in the system is diffusive decay associated with wavelengths of order $L$). Hence $\chi\sim L^2$ which is consistent with $\mathcal{X}=O(1)$ and $\chi\sim L^2$ (these results are in the regime $\lambda>\lambda_c$, see also (\ref{equ:def-chi}) and Figure~\ref{fig:AlgRes}).

The conclusions of this analysis are (i) that the longest relaxation time in the system controls the convergence with respect to $\tmax$ and (ii) that these relaxation times can be revealed by explicitly computing transients and autocorrelation functions as in Figs.~\ref{fig:obsAve},\ref{fig:AutoCorr}.   We also emphasize that making these measurements are not only useful for verifying convergence of the algorithm: they also reveal the important physical effects at work in the biased trajectories of interest: for $\lambda>\lambda_c$, the dominant physical effect is that the density becomes inhomogeneous on the macroscopic scale, so that $\langle |\delta\rho_1|^2\rangle_s$ diverges with system size.  There is a slow time scale associated with this macroscopic inhomogeneity, which scales as $\tau\sim L^2$ and results in a large susceptibility $\chi$.  This long time scale necessitates a large $\tmax$ in the cloning algorithm, since accurate estimates of observables like $\langle K\rangle_s$ require $\tmax\gg\tau$.

\subsection{Population Size $\Nc$ Convergence}
\label{subsec:ncConv}

\begin{figure*} 
\includegraphics{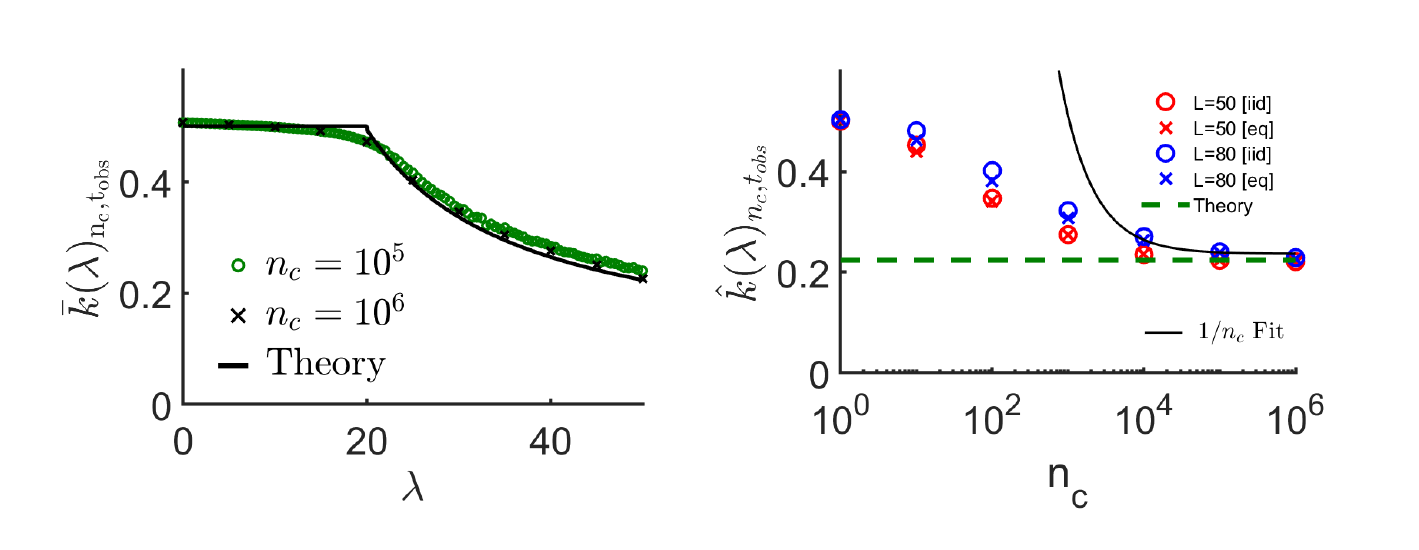}
\caption{Estimates $\overline{k}_L(\lambda)$ of the activity, as the number of clones is increased.  (a)~Data for $L=80$ and $\Nc=10^5,10^6$, as $\lambda$ is varied.  The theory line is the result for $L\to\infty$. (b)~Data for $\lambda=50$ and $L=50,80$ as $\Nc$ is varied. In all cases, $\tmax=10^4$ and $R=10$. 
The fit uses equation (\ref{equ:fit-finite-nc}) and is applied to the data for $L=80$ [eq], using only the points at $\Nc=10^4,10^5$.  This fit captures the scaling for large $\Nc$.  See also Table~\ref{table:efficiency-nc}.}
\label{fig:nc}
\end{figure*}

\begin{table}
\centering
\begin{tabular}{ c |  c | c | c | c |}
     & \multicolumn{2}{c} {L=50} \vline & \multicolumn{2}{c} {L=80}\\ 
     \cline{2-5}
     & [iid] & [eq] & [iid] & [eq] \\
     \hline
     Fit parameter $A$ (exc. $\Nc=10^6$) & 130 & 140 & 340 & 270\\
     Fit parameter  $k_\infty$ (exc. $\Nc=10^6$) &  0.2221 & 0.2220 & 0.2369 & 0.2367 \\
     Fit parameter $k_\infty$ (incl. $\Nc=10^6$) & 0.2214 & 0.2204 & 0.2322 & 0.2311\\
     Difference in $k_\infty$ & 0.3\% & 0.7\% & 2.0\% & 2.4\%  \\ 
\end{tabular}
\caption{Results of fitting the data in Fig.~\ref{fig:nc} to Equ~(\ref{equ:fit-finite-nc}).  This is analogous to the analysis of Table~\ref{table:efficiency-T} but the state point is different (so numerical values of $k_\infty$ are different).  The fits use data for $\Nc\geq10^4$.  The first two rows are results of fitting just two points $\Nc=10^4,10^5$.  The third row shows the estimate of $k_\infty$ when the final point at $\Nc=10^6$ is included in the fit.  For $L=80$, there is a small but systematic shift in the estimate of $k_\infty$ as more data is included, indicating that subleading corrections to scaling are not negligible at $\Nc=10^4$. }
\label{table:efficiency-nc}
\end{table}

This section shows the convergence of the algorithm as $\Nc$ increases.  In all cases we take $\tmax=10^4$, which is large enough that results depend weakly on $\tmax$.
Fig.~\ref{fig:nc} shows results, for a range of system sizes and numbers of clones.  From Fig.~\ref{fig:nc}a, one sees that for the relatively large system size $L=80$, a good estimate of $\KK$ is available already for $\Nc=10^5$, but increasing to $\Nc=10^6$ clones and focussing on $s>s^*$, the results still depend on $\Nc$, indicating that the large-$\Nc$ limit is not fully converged for $\Nc=10^5$.  (For $s<s^*$, the results do not agree perfectly with theoretical prediction shown in Fig.~\ref{fig:nc}a: this is because the theory applies only in the large-$L$ limit, but these are finite systems.  The important feature for convergence is not the agreement with the analytic theory, but whether the results depend significantly on $\Nc$.)  Fig.~\ref{fig:nc}b shows results at the state point $\lambda=sL^2=50$, which is in the phase-separated regime (recall Fig.~\ref{fig:Traj}): one sees that the number of clones required to obtain accurate results is of order $10^5-10^6$, with the larger system requiring larger $\Nc$.

To investigate this in more detail, Fig~\ref{fig:nc} shows a fitting analysis similar to that of Fig.~\ref{fig:T}, with a fit function
\begin{equation}
\bar{k}(\lambda)_{n_c,\tmax} = k_\infty ( 1 + A/\Nc ) + O(\Nc^{-2}) ,
\label{equ:fit-finite-nc}
\end{equation}
 analogous to (\ref{equ:fit-finite-tmax}). 
 This is the large-$\Nc$ scaling form predicted by~\cite{nemoto17finite,hidalgo17finite}.  As was the case for large $\tmax$, a reasonably accurate extrapolation to the last data point ($\Nc=10^6$) can be obtained by a fit through the two previous data points ($\Nc=10^4,10^5)$.  However, the quantitative analysis of Table~\ref{table:efficiency-nc} shows that for $L=80$ there is a significant ($\sim 2\%$) shift in the estimated value of $k_\infty$ when the data from $\Nc=10^6$ are included. This indicates that there are systematic deviations from the fit, presumably because data for $\Nc=10^4$ are not yet in the asymptotic regime described by (\ref{equ:fit-finite-nc}).


\begin{figure*} [t]
\includegraphics{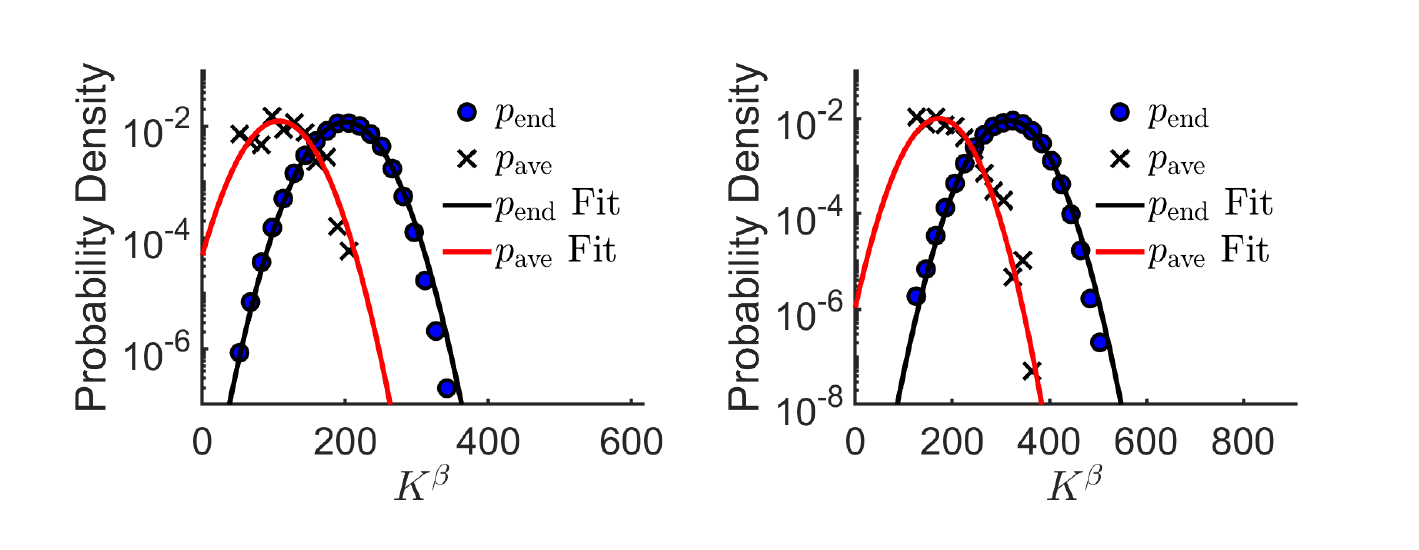}
\caption{Distributions of the activity $K^\beta(\Theta_i)$ at $\beta=970$ when $\tmax=10^4$ and $\Delta t=10$. (a)~$L=50$; (b)~$L=80$. The data points are measured across $\Nc=10^6$ systems and Gaussian fits produced by measuring the average $\mu$ and variance $\sigma^2$ across the population. All results are at $\lambda = 50$.}
\label{fig:KDist}
\end{figure*}

To understand why these large numbers of clones are required, it is useful to return to the distributions $\pave$ and $\pend$ defined in (\ref{eq:Opave},\ref{eq:Opend}).  We construct these distributions for the observable $K^\beta$, the number of particle hopping events in the time interval $[t_\beta,t_{\beta+1}]$.   Results are shown in Figure~\ref{fig:KDist}.  Recall that the distribution $\pend$ is representative of the transient regime ($t\approx\tmax$), while the distribution $\pave$ is representative of the time-translation-invariant (TTI) regime.  Accurate characterisation of the TTI regime is essential for obtaining  accurate estimates of $\KK_L$.

However, we note that the operation of the cloning algorithm means that the distribution $\pend$ is sampled directly, but $\pave$ is obtained by  a form of importance sampling from the distribution $\pend$.  In terms of Figure~\ref{fig:KDist}, this means that data is only available for $\pave$ over a restricted range, which is the range over which data is available for $\pend$.  If these two distributions have a low overlap, then the algorithm will not sample $\pave$ correctly: this is an important source of the systematic errors in Figure~\ref{fig:nc}.  In fact, this argument does not only apply to the distributions $\pend$ and $\pave$ of the order parameter $K^\beta$ -- one requires that for any observable (say $F$) the distributions $\pave(F)$ and $\pend(F)$ should overlap, otherwise the values of $F$ sampled within the algorithm will be biased, leading (potentially) to systematic errors.  The possible choices of observable include variants of $K^\beta$ in which the averaging interval $[t_{\beta-1},t_\beta]$ is replaced by some other interval, eg $[t_\beta-t_0,t_\beta]$: the overlap of $\pave$ and $\pend$ for this distribution will likely have a non-trivial dependence on $t_0$.  In this sense, significant overlap of the distributions shown in Fig.~\ref{fig:KDist} is a necessary but not a sufficient condition for convergence: one requires in general that these distributions overlap for all observables $F$ (otherwise the sampling of states within the TTI regime will be biased away from its correct distribution).

To estimate the significance of this effect (see also~\cite{hurtado09}),  we focus for convenience on the distribution of $K^\beta$ and suppose that both $\pave$ and $\pend$ are described by (approximately) Gaussian distributions with mean values $\mu_{\rm ave}, \mu_{\rm end}$ and the same variance $\sigma^2$.  (Note $\mu_{\rm ave}=\mathcal{K}_L$.) To be precise, define $g(x,\mu,\sigma)=\exp(-(x-\mu)^2/2\sigma^2)/\sqrt{2\pi\sigma^2}$ and suppose $\pend(K)\approx g(K,\mu_{\rm end},\sigma)$, and similarly for $\pave$.  Also define a scaled version of the complementary error function as $G_\sigma(x)=\int_x^\infty g(y,0,\sigma)\mathrm{d}y$.  Given $\Nc$ clones drawn independently from $\pend$, we expect to sample a range of $K$-values $(\mu_{\rm end}-A) < K < (\mu_{\rm end}+A)$ with $G_\sigma(A)=(1/\Nc)$.  We further assume that the measured $\pave$ is close to the true $\pave$ over the range over which data is available -- this is certainly an approximation but it allows an estimate of the effect of the unsampled range on the results of the algorithm. With this approximation, the average value of $K$ with respect to $\pave$ can be estimated as
\begin{equation}
\overline{k} \approx \frac{ \int_{\mu_{\rm end}-A}^{\mu_{\rm end}+A} K g(K,\mu_{\rm ave},\sigma) \mathrm{d}K }
     {  \int_{\mu_{\rm end}-A}^{\mu_{\rm end}+A} g(K,\mu_{\rm ave},\sigma) \mathrm{d}K}
\label{equ:kbar-est}
\end{equation}
Assuming as in Figure~\ref{fig:KDist} that $\mu_{\rm end}>\mu_{\rm ave}$ (and $\sigma$ is not too large), one may replace the upper limits in (\ref{equ:kbar-est}) by infinity. 
Writing $\Delta\mu=(\mu_{\rm end}-\mu_{\rm ave})$ , this yields 
\begin{equation}
(\overline{k} -  \mu_{\rm ave} ) \approx \sigma  / \sqrt{2\pi} \cdot \frac{\exp( - (A-\Delta \mu)^2 / 2\sigma^2 )}{G_\sigma(\Delta\mu-A)}
\label{equ:kerr}
\end{equation}
This error converges to zero as $A\to+\infty$, as it should do.  However, for large $\Nc$ one has the scaling relation $A\sim \sigma\sqrt{\log \Nc}$ so $A$ grows slowly with $\Nc$.   The relevant dimensionless parameter for convergence is $X=(A-\Delta\mu)/\sigma$: this parameter is positive if the peak of the $\pave$ distribution is within the range of the sampled data: accurate results require $X\gtrsim1$.
This requires
\begin{equation}
\Nc \gtrsim \exp\left[ (\Delta\mu/\sigma)^2 \right] .
\label{equ:nc-scaling}
\end{equation}
In general, one expects $\Delta\mu$ to increase as the biasing field increases; on the other hand $\sigma$ is expected to depend weakly on the bias but (since $K$ is extensive in space) one expects $\sigma\sim1/L$ in large systems.  Hence one expects that the number of clones required to achieve convergence increases exponentially in the system size and in the distance from equilibrium.  For a similar argument, see~\cite[Section 5]{hurtado09}.  Based on the Gaussian fits in Figure~\ref{fig:KDist} the lower bounds on $\Nc$ are $2.16 \times 10^3$ ($L=50$) and $4.93 \times 10^4$ ($L=80$): for high-accuracy estimates at these state points, we require $\Nc=10^5$ and $\Nc=10^6$ respectively, so high accuracy requires parameters significantly in excess of these lower bounds, consistent with this argument.

\section{Computation}
\label{sec:comput}

In this section, we discuss the performance of our computational implementation of the cloning algorithm.  Given the large number of clones evolving independently, it is natural to use high-performance (parallel) computing methods to speed up the computations.  However, the step where clones are copied and deleted involves a significant communication overhead, which can substantially reduce efficiency. Close to the phase transition ($\lambda=\lambda_c\approx20$) the variance in activity is close to its maximum. This suggests that there should be a large variance in the $\Upsilon^\beta(\Theta_i)$, and a large amount of cloning activity. This makes the phase transition a useful place to test the algorithm.

We compare results obtained using MPI with simple serial computations and a shared-memory (OpenMP) implementation. We define the speed-up factor of an implementation $m$ as 
\begin{equation}
\mathcal{S}_m = \text{RT}_s/\text{RT}_m
\end{equation} 
where $\text{RT}_m$ is the run time of implementation $m$ and $\text{RT}_s$ is the run time for a simple serial code.  For a computation that uses $n_{\rm t}$ threads, we define the efficiency as
\begin{equation}
\mathcal{E}_m = \mathcal{S}_m/n_{\rm t}
\end{equation} 
Both the speed-up and the efficiency are depend in general on $n_{\rm t}$.  We describe an OpenMP implementation that is 99\% efficient (but does not scale beyond a single computational node). For performance on multiple nodes, we use a method based on the message-passing interface (MPI) protocol, which achieves an efficiency of around 90\% on computations distributed over 64 processors.  We measure weak scaling performance up to 128 processors. All codes were written in C++ and will be available shortly after publication. 

All results were obtained on (8-core) Intel Xeon E5-2650V2 Ivybridge processors running at 2.6 GHz.   As a representative test case, we consider a system of size $L=50$ at parameters at which the activity estimator $\fs$ has converged, $\tmax=10^4$ and $\Nc=10^5$. We average all results over 10 independent computations.   This computation takes 5500 seconds to run on a single core.  We present results for efficiency relative to this reference point.  The shared memory implementation uses OpenMP with 16 threads on a dual-socket node.  For the MPI implementation, we typically use four such nodes connected using Infiniband QDR with a total of 64 threads.

\begin{figure}[t]
\centering
\includegraphics{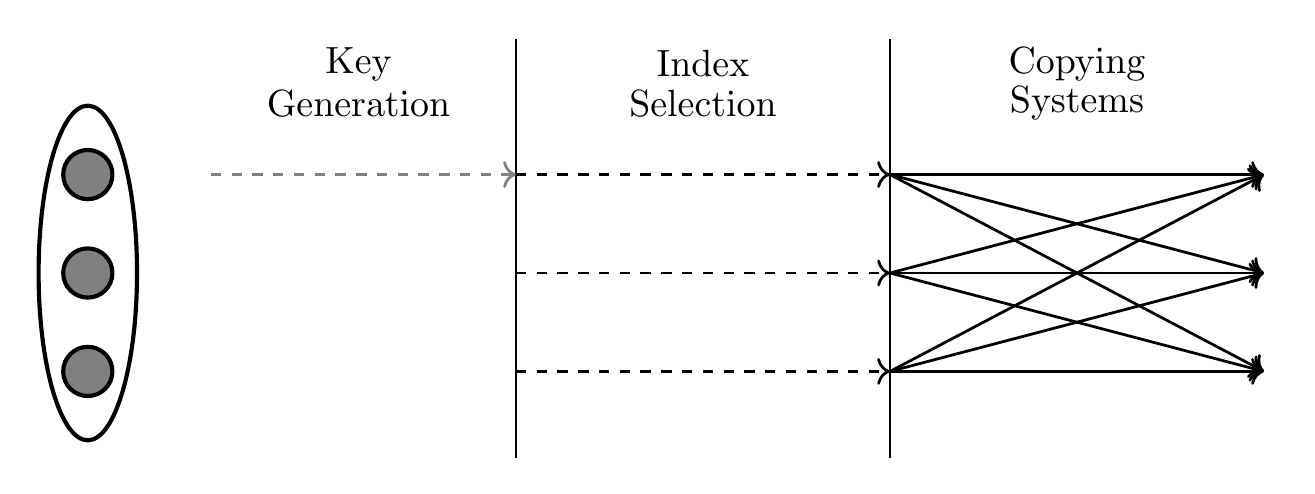}
\caption{OpenMP implementation of the algorithm code run on one node (ellipse) and multiple threads (grey circles).}
\label{fig:OpenMP Implementation}
\end{figure}

\subsection{Serial and OpenMP Implementations}

The simplest implementation is a serial code, which simulates the dynamics for each clone in turn.  The cloning step involves copying and deletion of the different clones.
This is achieved by each clone in the new population ``pulling'' its state (that is, the site occupancies, particle positions, etc) from a member of the old population~\footnote{The new and old population are stored separately in memory, so every cloning step involves $\Nc$ copy operations.  
If the new and old population are similar to each other, some of these copy operations could be avoided by updating the original population (in-place) instead of copying the new population from the old one.  While this might improve efficiency in some cases, the benefits are
negligible for the values of $\Delta t$ considered here, since the new and old populations differ substantially, and the main computational effort comes from the system dynamics, not the copying.  The advantage of copying the population instead of updating (in-place) is that it simplifies the implementation of the algorithm, since the copy operations are completely independent and happen in parallel.}.

The choice of which member of the old population is copied into clone $j$ of the new population is specified by the number $\alpha_j$ (see Sec.~\ref{sec:alpha}). One then has to find the value $k$ that solves (\ref{equ:alpha-Ups}): this is achieved by a binary search.  The \emph{key} for this search is an ordered list of real numbers, in which the $k$th element is $\sum_{i=1}^k \Upsilon^\beta(\Theta_i)$.  
Random number generation uses the Mersenne Twister algorithm~\cite{mersenne}, with one instance of the random number generator for each clone.  Except where otherwise stated (see below), this means that the results are fully reproducible, independent of whether the serial or parallel code is used.

In the OpenMP (shared memory) implementation, the clones are distributed equally over 16 threads on a single computational node. The dynamics are simulated as in the serial case. 
In the cloning step, the binary search key is constructed using a single thread, but the copying of clone states is done in parallel. A diagram summarising this process is shown in Figure~\ref{fig:OpenMP Implementation}.  For our test case ($L=50$, $\tmax=10^4$, $\Nc=10^5$, $\lambda=20$), the speed-up of the OpenMP implementation with respect to the serial code is 15.8, corresponding to an efficiency of 99\%. 
Efficiencies at other state points are similar.  

\begin{table}
\centering
\begin{tabular}{ c |  c | c}
     
     Method & Speed-Up & Efficiency (\%)\\
     \hline
  
     One clone per message & 48.3 & 75.5 \\
     Many clones per message & 55.2 & 86.2\\
     Reduced communication & 57.2 & 89.3\\ 
\end{tabular}
\caption{Speed-ups and efficiencies relative to our serial implementation obtained by MPI implementations when using the equal spacing [eq] clone selection method with different amounts of packing when running on 64 processors distributed across 4 nodes. The results are obtained at $\lambda=20$ and $L=50$ with $\tmax=10^4, \Nc=10^5$ and $\Delta t=10$.}
\label{table:efficiency}
\end{table}

\begin{figure} [t]
\centering
\includegraphics{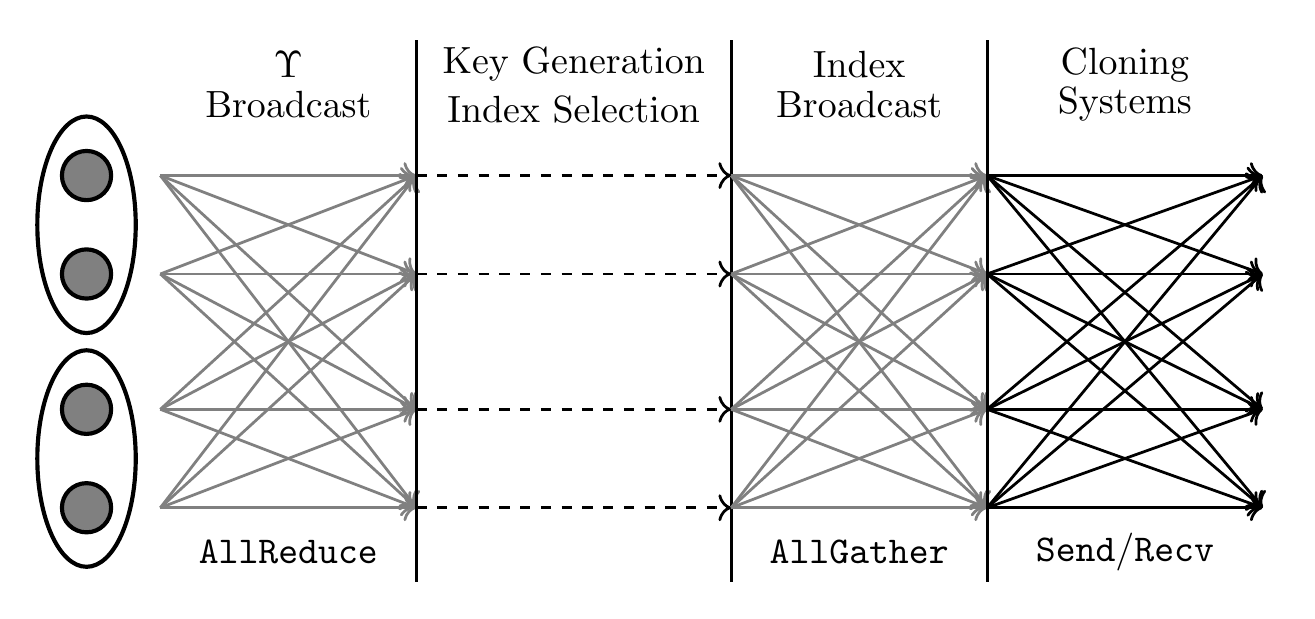}
\caption{MPI implementation of the algorithm code run on two nodes (ellipses) and multiple threads (grey circles). Arrows are MPI communications, black arrows are pointwise and grey arrows are collective.}
\label{fig:MPI Implementation}
\end{figure}

\subsection{MPI Implementation}

In order to exploit higher levels of parallelisation, we also use a distributed memory (MPI) implementation, which is illustrated in Figure~\ref{fig:MPI Implementation}.  The direct generalisation of the OpenMP code gives rise to an implementation that we call ``simple communication''.  In this case, the MPI method produces identical output to the OpenMP method.  We also describe a method with reduced communication, in which case the results are statistically equivalent (but not identical), because the systems being simulated are distributed differently among the relevant threads, and hence the instance of the random number generator that is used to simulate a particular system is different.

\subsubsection{Simple communication.}

In the dynamical stage of the algorithm, the clones are distributed evenly across the threads.  In the cloning step, each thread sends the $\Upsilon^\beta(\Theta_i)$ for its clones to all other threads (this is an MPI {\tt AllReduce} communication).  Each thread constructs the binary key independently (this computation is therefore duplicated for each thread, but this is  more efficient than calculating it on one thread and broadcasting it to all others).  With the binary key in place, each clone in the new population uses (\ref{equ:alpha-Ups}) to decide from which  clone to pull its state (the value of $d$ in (\ref{equ:alpha-Ups}) is produced by thread 0 and broadcast to all threads after the key generation).  Next, the indices of the clones that are required by each thread are communicated to all other threads (MPI {\tt AllGather} operation); based on that information, the clone states are sent between the nodes, as required (using many MPI {\tt Send} operations).  In the simplest case, the state of each clone is sent using a single MPI message (we call this ``one-clone-per-message'').  Alternatively, all information to be exchanged between each pair of threads can be packed into a single message (``many-clones-per-message'').

From Table~\ref{table:efficiency}, these two methods achieve efficiencies of 75-86\% on systems of $n_{\rm t}=64$ threads, with the many-clones-per-message method being more efficient.  We emphasise that the amount of information being exchanged between each thread is identical in this case, so the increased efficiency comes from packing the same information into a smaller number of messages.  For one-clone-per-message, the total number of messages is $\Nc=10^5$; for many-clones-per-message, each thread receives at most one other message from each other thread, so the total number of messages is at most $n_{\rm t}(n_{\rm t}-1)=4032$ (see Sec.~\ref{sec:comm}, below).  This is a reduction by more than an order of magnitude in the number of messages, which significantly improves performance. 

\subsubsection{Reduced communication.}

The simple communication method is somewhat inefficient because in some situations, a particular thread ($A$) may send several clones to  some other thread ($B$), but thread $B$ also sends several clones back to thread $A$.  To avoid this redundancy, we use a cancellation procedure where each thread preferentially pulls states from clones that are on the same thread.  (This is achieved by pairwise cancellation of clones that have been designated by (\ref{equ:alpha-Ups}) to be sent between threads; the resulting new population is identical but its partitioning among the threads is different.)  With this ``reduced communication'' method, clones are sent either from $A$ to $B$ or from $B$ to $A$ (instead of sending one message in each direction).  This typically reduces the number of messages by almost a factor of 2, and the size of the resulting messages by a factor of 9.  Table~\ref{table:efficiency} shows that this leads to an improvement in efficiency.

\subsubsection{Communication Patterns}
\label{sec:comm}

\begin{figure} [t]
\centering
\includegraphics{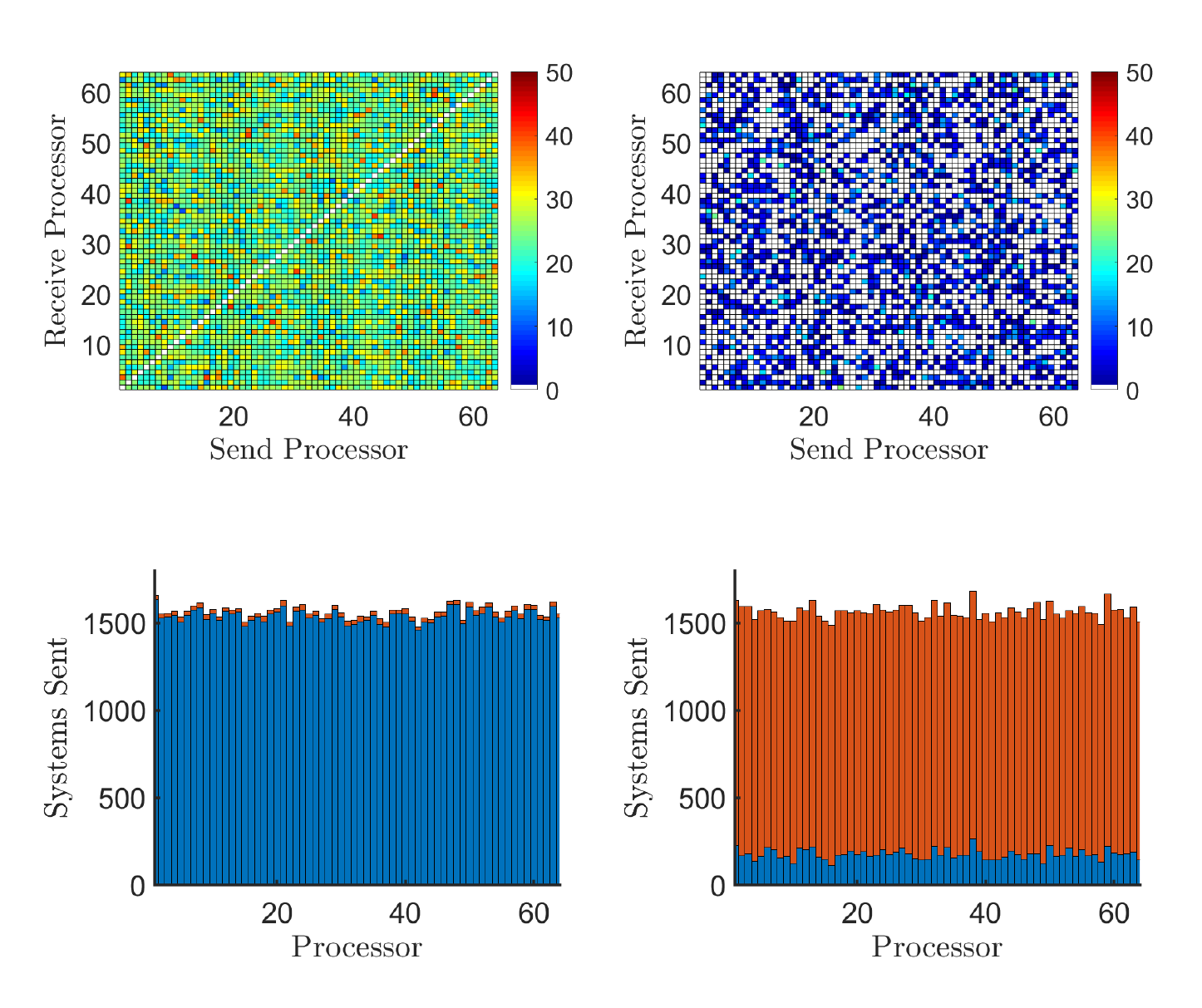}
\caption{The number of systems sent between each pair of processors using MPI when evolving $\Nc=10^5, \tmax=10^4$ and $L=50$. Simple communications (\textit{left}) and reduced communications (\textit{right}) using the independent [iid] clone selection method. These communications are in the $500^{th}$ cloning interval  ($t=5000$) and the cloning intervals are $dt=10$ units of time. The algorithm runs at a bias $\lambda = 20$. \textit{Bottom Row:} The number of systems sent by each processor using MPI (blue) and those whose states are simply copied (orange).}
\label{fig:CommsPatterns}
\end{figure}

To illustrate the communications pattern, Figure~\ref{fig:CommsPatterns} shows the number of messages sent and received by each thread in a typical cloning step at $t = \tmax / 2 = 5000$. In the simple implementation, each thread typically sends around 30 clones to each other thread and receives a similar number.  In the reduced communication implementation, there are less than half as many MPI communications and those that occur contain about 5 systems.

\subsubsection{Weak Scaling}

For a fixed number of threads $n_{\rm t}$, the run time for each algorithm scales linearly with the number of clones $\Nc$. A \emph{weak scaling} analysis is one in which the problem size (in this case $\Nc$) is increased, with a proportionate increase in the computational resource (the number of nodes $n_N=n_{\rm t}/16$).  In the best (most efficient) case, this leads to a run-time that remains constant as the problem-size $\Nc$ increases.  In general, one expects the run-time to increase with problem size, with a smaller increase corresponding to a more efficient algorithm.

\begin{figure} [t]
\centering
\includegraphics{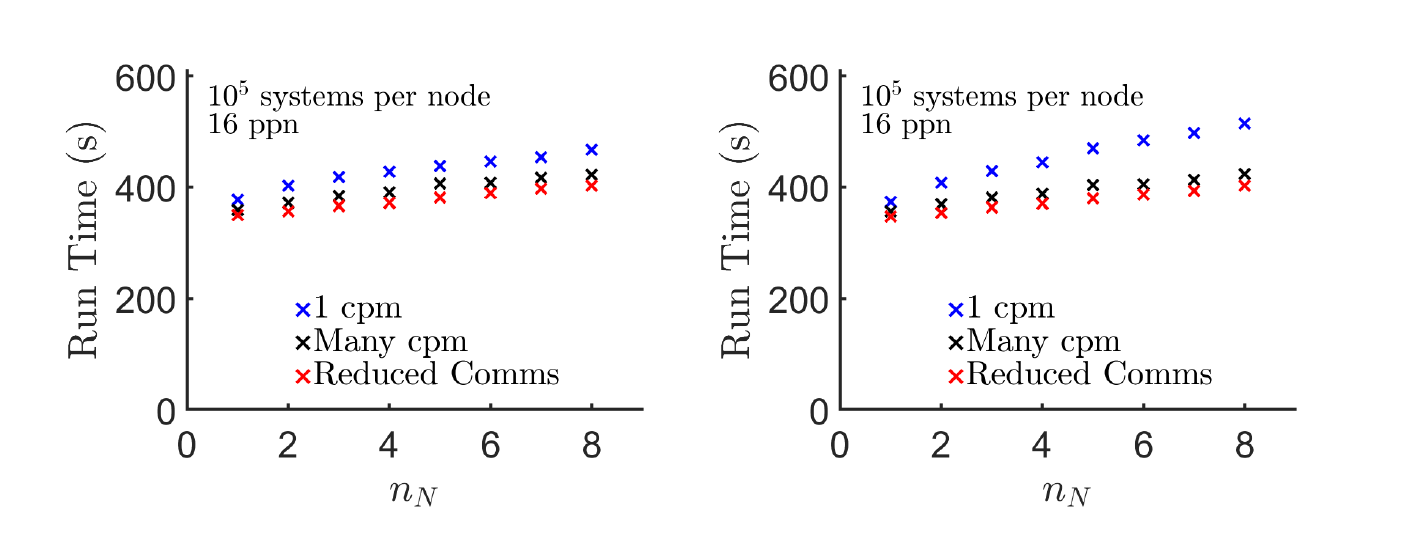}
\caption{The weak scaling of the run time of the code for both clone selection methods [iid] (left) and [eq] (right). We fix $\Nc/n_N=10^5$ and the code is run at $\tmax=10^4$, $L=50$, $\lambda=20$ and $dt=10$. We vary the number of clones per message (cpm) and reduce the communications.}
\label{fig:WScaling}
\end{figure}

In Figure \ref{fig:WScaling} we measure the weak-scaling of a system at $L=50$ with enough clones ($nc/n_N=10^5$) that the algorithm is converged when the code is run on one node. The value of the bias ($\lambda=20$) that we consider is very close to the phase transition and the value of the estimator $\fs$ depends only very weakly on $\Nc$. We show results for both the [eq] and [iid] clone selection methods, but the results are similar in both cases.

As the problem size increases, one observes a gradual increase in run time: for the most efficient algorithm, this increases by approximately 5\% while the number of clones has increased by a factor of 8.  This shows that parallel-computing platforms significantly reduce the wall-time required to perform large computations, although (as expected) the method is most efficient when run on just a few nodes, since the communication overhead is lower in that case.  The gradual increase in run-time with $n_N$ does not indicate any optimum value for $n_N$ in this case nor any cutoff beyond which the method becomes inefficient: the code is performing well across this range of $n_N$ (which corresponds to at most 128 parallel threads).  

\label{subsec:WScal}

\section{Conclusion}
\label{sec:conc}

We have analysed the finite-size scaling of the dynamical phase transition in the SSEP.  Our results show that the cloning algorithm can demonstrate convergence of the limit of infinite system-size $L \to \infty$ in this problem, although of order $10^6$ clones are required to achieve this. By analysis of the $\pave$ and $\pend$ distributions of the activity we can understand the number of clones required for convergence. We also analysed the time $\tmax$ required for convergence of the limit $\tmax\to\infty$, by measuring of the transient decay of several observables, and their associated autocorrelation functions.

We have achieved computational efficiencies of 99\% using an OpenMP implementation of the algorithm, although this requires shared memory for all threads and so can be used on a single node. We also used an MPI implementation, so that the algorithm can be scaled across multiple nodes. Optimising MPI communication to use as few messages as possible significantly increases the algorithmic performance. We have achieved this by point-wise reduction of the MPI communications between pairs of processors and packing multiple systems into each MPI message. By measuring the weak-scaling of run time across multiple processors we find that there is a slow drop off in efficiency. 

The implementation of the cloning stage of the algorithm can affect the errors in the method. We have made a comparison of two methods for selecting which systems are cloned, one of these methods reduces the number of systems that are deleted. Whilst both methods produce similar systematic errors and require similar computation time, we have found that the method which reduces the number of system deletions incurs much smaller statistical errors.

As the cloning algorithm becomes more widely used~\cite{nemoto2017,limmer2017importance,espigares17}, we hope that these methods for error analysis and parallel computing implementation will be useful for the growing effort into research on large deviations of time-averaged quantities.

\ack RLJ thanks Takahiro Nemoto and Vivien Lecomte for useful discussions about the population dynamics algorithm. We also thank the anonymous referees for insightful comments and suggestions.  Computations were performed on the HPC service at the University of Bath.

\section{References}

\bibliographystyle{iopart-num}  
\bibliography{MyBib,extra}

\end{document}